\documentclass[11pt]{article}
\usepackage{amsmath,tabu,array}
\usepackage{epsfig}
\usepackage{graphicx}
\usepackage{color}
\usepackage[normalem] {ulem}
\usepackage{soul}

\usepackage{verbatim}
\definecolor{violet}{rgb}{0.5,0,0.5}
\definecolor{gris25}{gray}{0.5}
\definecolor{edo}{rgb}{0.0,0.5,0.5}
\definecolor{edo2}{rgb}{1.0,0.5,0.5}
\definecolor{edo3}{rgb}{0.8,0.1,0.8}

\begin{document}

\title{\vspace{-3cm} Numerical study of DNA denaturation with self-avoidance: 
pseudo-critical temperatures and finite size behaviour.}

\author{Barbara Coluzzi$^{a,}$
\footnote{Corresponding author. E-mail: \tt barbara.coluzzi@sbai.uniroma1.it}
\hspace{.2cm} and Edouard Yeramian$^{b}$}

\maketitle

\begin{center}

a) {\em Dipartimento di Scienze di Base ed Applicate per l'Ingegneria}
(RM002 - room 24), 
\\Sapienza University, via A. Scarpa 14, 00161 Rome, Italy.

b) {\em Unit\'e de Microbiologie
Structurale}, UMR 3528 CNRS, Institut Pasteur, \\ 25-28 rue du
Docteur Roux, 75724 Paris cedex 15, France\\

\begin{abstract}
\noindent
We perform an extensive numerical study of the disordered 
Poland-Scheraga (PS) model for DNA denaturation in which self-avoidance is 
completely taken into account. In complement to our previous work, 
we focus here on the finite size scaling in terms of pseudo-critical 
temperatures. We find notably that the mean value and the fluctuations of the 
pseudo-$T_c$ scale with the same exponent, the correlation length exponent 
$\nu_r$ (for which we provide the refined evaluation 
$\nu_r=2.9 \pm 0.4$). This result (coherent with the 
typical picture that describes 
random ferromagnets, when disorder is relevant) is at 
variance with numerical results reported in the literature for the PS model 
with self-avoidance, 
leading to an alternative scenario with a 
pseudo first order transition. 
We moreover introduce a crossover chain length $N^*$, which we evaluate, 
appropriate 
for characterizing the approach to the asymptotic regime in this model.
Essentially, below $N^*$, the behaviour of the model in our study
could also agree with such alternative scenario. 
Based on an approximate 
prediction of the dependence of $N^*$ on the parameters of the model,
we show that following the choice of such parameters it could be not possible to
reach the asymptotic regime in practice. In such context it becomes then possible to reconcile 
the apparently contradictory numerical studies.

\end{abstract}
\end{center}

\noindent
PACS: 64.60.Fr \hspace{.15cm} Equilibrium properties near critical points, 
critical exponents \\
PACS: {82.39.Pj}\hspace{.35cm}{Nucleic acids, DNA and RNA bases} \\
PACS: {89.75.Fb}\hspace{.3cm}{Structures and organization in complex systems}

\newpage

\topmargin=-1.5cm

\section{Introduction}
\noindent
The study of disordered systems has attracted a wide interest in 
statistical physics. Numerical 
studies appear of critical importance 
in the field, as the handling of disorder can be difficultly amenable to 
analytic treatments. In any case numerical approaches appear to be the best 
suited for
capturing details of interesting 
behaviours, such as those related to finite-size effects.

In this background, the disordered Poland-Scheraga (PS) model 
for DNA denaturation stands as a privileged model, 
following several perspectives: i) as an instance of the so-called "helix-coil 
transition" models,
the model is designed to capture one of the most basic phenomenon in
molecular biology, relative to the opening of the double-helix; ii) to render 
the helix-coil transition model more realistic, the PS model
takes into account intrinsic long-range effects (associated with the weights 
of the loops in the coiled state), which can extend throughout
the length of the system; iii) originally specifically 
associated with the model, 
numerical methods were developed to handle the statistical mechanics problem 
with long-range effects
very efficiently, allowing to consider system sizes which are orders of 
magnitude larger than what would be possible with Monte-Carlo methods.

With this respect, the PS model
of DNA denaturation with self-avoidance is of particular interest, 
notably as it belongs to
a more general class of polymer models for which the question of
disorder relevance has been definitely 
answered only recently \cite{GiTo1,GiTo2,Gi}, in a probabilistic mathematical 
framework.
More precisely, for this class of models, disorder is predicted to be 
relevant, the transition being 
expected to be at least of second order. 
In such context, the numerical findings of our previous 
work \cite{CoYe}, relying on a standard finite size scaling analysis, indeed 
agree with a smooth transition, suggesting a value $\nu_r=2.9 \pm 0.6$ for 
the correlation length exponent.

In contrast, other studies of the DNA denaturation model with self-avoidance
\cite{GaMo,MoGa} instead supported, both on numerical and 
theoretical grounds, the possibility of 
a pseudo first order transition. 
Such scenario could be in fact accommodated
within theoretical descriptions of relevant models, such 
as random ferromagnets, with, in any case, disorder expected to be 
relevant from the Harris criterion point of view. 
Briefly, within this 
picture, one predicts the presence of two correlation lengths, which can be 
evidenced numerically in particular by an analysis in terms of 
(appropriately defined, sequence-dependent) {\em pseudo-critical 
temperatures}, since the two corresponding exponents should rule the 
scaling of the mean value and that of the fluctuations of the pseudo-$T_c$, 
respectively. 
In detail, in these studies, the exponent 
related to the fluctuations was found to have the value 2, thus coherent with 
a second order transition (and, from this point of view, with
disorder relevance), whereas the one related to the mean value was found 
to have the value 1, thus, concomitantly, denoting a pseudo first 
order character of this transition.

In fact, based on recent mathematical findings, 
\cite{GiTo1,GiTo2,Gi} it would be possible to rule out such 
scenario for the thermodynamic limit behaviour of the 
present model. This situation is further reinforced by the prediction,
within the same probabilistic 
mathematical framework, of a thermodynamic limit behaviour governed 
by a single correlation length \cite{To}.

The general situation as described above motivates the present extensive 
numerical study, in the  prolongation of our previous work, since 
it appears desirable to clarify the overall situation relative to the 
theoretical/numerical 
results, all the more that in \cite{GaMo,MoGa} extremely large sizes 
were considered (up to $N = 2 \cdot 10^6$), 
which should in principle be appropriate for reflecting the thermodynamic 
limit properties evidenced in the theoretical studies. 
It is in this direction that we perform here a finite size scaling analysis in 
terms of pseudo-critical temperatures. In any event,
for the present peculiar model, 
displaying strong corrections to 
scaling and a slow approach to the asymptotic regime \cite{CoYe}, 
a general 
deeper understanding of the finite size behaviour is 
of interest both from the physics and biological points of view.

 More precisely, 
we tackle a hitherto unaddressed problem, relative to the definition of 
appropriate pseudo-critical temperatures in the presence of multiple peaks in 
quantities such as the specific heat and the susceptibility, which represent 
as a matter of fact a salient feature of the underlying biological model.
Focusing on the strongly 
non self-averaging behaviour of observables at the critical point, 
we find that the pseudo-$T_c$ can be appropriately defined as 
the position of the absolute maximum of susceptibility, for a given sequence. 
We notably show that the mean value and the fluctuations of this observable
scale with the same exponent, thereby confirming that the
asymptotic regime is reached in our case. At variance with the numerical 
findings
in \cite{GaMo,MoGa}, the present result is thus coherent with the 
typical picture describing random ferromagnets 
when disorder is relevant. 
  
In order to understand in depth the 
finite size behaviour, we further perform an extensive analysis
of the loop-length probability distribution, at different temperatures.
The behaviour of this observable reveals in a particularly evident way the 
presence
of a {\em crossover} chain length $N^*$. 
Indeed, this 
quantity appears to be particularly appropriate for characterizing the slow 
approach to the asymptotic regime in the present model. Basically, for chain 
lengths below $N^*$, in our study the behaviour of the model would also agree 
with the 
scenario of a pseudo first order transition. 
As a matter of fact, the evaluation 
of $N^*$ in our case, and the approximate prediction within a phenomenological 
framework of its dependence on the parameters of the model,  
allows us to resolve the possible paradox concerning contradictory
numerical findings as described above. We show notably that, with the 
choice of parameters in \cite{GaMo,MoGa}, it could be not possible to
reach the asymptotic regime in practice.

The paper is organized as follows: we briefly review the general background 
in Section \ref{zero}; 
we introduce the model and the observables in Section \ref{beginning}, 
where we first summarize known results for the pure case 
(\ref{beginning_one}), we introduce the disordered model, by defining the 
different parameters entering in the computation of the partition 
function by means of the recursive equations (\ref{beginning_two}), we 
describe possible scenarios 
within the framework of the analysis in terms of pseudo-critical 
temperatures, clarifying the 
different ways of averaging considered (\ref{beginning_three}), we recall the 
approximation used for the power law and the details of the numerical 
implementation of the computations (\ref{beginning_four}); we present and 
discuss our results on the analysis with the 
pseudo-critical temperatures in Section \ref{discussion1}, where we start 
from the possible definitions of 
this observable (\ref{discussion1_one}), 
we study the scaling of its mean value and of its 
fluctuations (\ref{discussion1_two}), we compare different ways of averaging 
(\ref{discussion1_three}), and we show data on the non self-averageness 
parameter related to the susceptibility (\ref{discussion1_four});
we present and discuss our results on the characterization
of the finite size behaviour in Section  \ref{discussion2}, where we 
outline the presence of the crossover chain 
length $N^*$ by the detailed study of 
data on the loop-length 
probability distribution 
(\ref{discussion2_one}), and we discuss 
our attempt to roughly estimate the dependence of 
this quantity on the parameters of the model (\ref{discussion2_two}); 
finally, we present our conclusions in Section \ref{ending}.

\section{General background}
\label{zero}
\noindent
A realistic model for the DNA denaturation transition (taking into account the 
entropic weights of the loops) was introduced by Poland and Scheraga 
(PS) \cite{PoSh1,PoSh2,Fi1}, to account for experimental melting curves,  
allowing their prediction \cite{review,Bl,JoEv}. Models for DNA denaturation 
were originally developed in the 
context of fundamental biological questions, relevant notably to the possible 
overlap between physical (helix/coil) and genetic (coding/non-coding) 
segmentations of genomic sequences (for an overview see 
\cite{Ye,YeJo,YeBoLa,CaMaBl}, and references therein).  
Furthermore, the PS model 
appeared to be 
also of interest in the context of statistical mechanics: the segmentation of 
the DNA chain into helix/coil regions represents an instance of an {\em almost 
uni-dimensional} system \cite{Fi0}, which can undergo a phase transition 
because of the long range effects, associated with the entropic weights of 
the loops (with the loop-length probability distribution 
described by an exponent $c_p$).

In fact, upon introducing the $c_p$ value appropriate for taking into account 
loop self-avoidance effects, larger than the one describing a random walk 
loop, it 
was hypothesized early on \cite{Fi1} that a more complete handling of
self-avoidance, taking also into account the self-avoidance
of the chain with itself, would lead to an even larger value for the exponent, 
and  correspondingly to a sharper transition in the $pure$ case (homogeneous 
sequence). However the confirmation of this hypothesis \cite{KaMuPe1,KaMuPe2} 
had to await conformal field theory results \cite{Du}, showing that the 
correct loop-length probability distribution exponent associated to a 
self-avoiding loop embedded in a self-avoiding chain in three 
dimensions fulfills the condition $c_p > 2$, leading to a 
first order singularity (with corresponding correlation length critical 
exponent $\nu_p=1$). The first order character of 
the transition was 
initially observed in a numerical study of a  
on-lattice homogeneous model consisting of two interacting self-avoiding walks 
(SAWs) 
\cite{CaCoGr}, with the numerical estimation 
($c_p \simeq 2.15$) in this case \cite{CaOrSt,BaCaSt,BaCaKaMuOrSt} 
in very good agreement with the theoretical predictions mentioned above. 

It can be noted that, because of the difference between the link energies 
of $\varepsilon_{AT}$ (Adenine-Thymine) and $\varepsilon_{GC}$ 
(Guanine-Cytosine) base pairs, DNA is an intrinsically $random$ system in which 
the disordered variables are $quenched$, in the sense that the composition of 
the sequence in terms of base pairs does not vary during the 
denaturation process. Accordingly, to understand the behaviour of the system in 
the thermodynamic limit of chain length $N \rightarrow \infty$ in the 
{\em canonical} ensemble, it is necessary to evaluate the quenched free energy 
density, by taking the average (over the random variables) of the logarithm 
of the partition function for the generated disordered configurations. However, 
it appears particularly hard to handle such quantity analytically \cite{KuLi}. 
In fact, whereas the equilibrium properties of the pure system 
were well understood, it was only recently that probabilistic 
mathematical approaches \cite{GiTo1,GiTo2,Gi,To} allowed to definitely 
demonstrate
that general theoretical results relative to the 
relevance of disorder \cite{Ha,ChChFiSp,AiWe,Be} do also hold for the depinning 
transition of disordered copolymers. This conclusion holds in particular for 
the PS models with $c_p>2$, despite the fact that such models are instances 
of peculiar first order transitions, characterized by the presence of a 
diverging correlation length, in 
a system in which the disorder couples to only one of the two 
phases. Accordingly, such cases are expected to exhibit a random fixed point 
corresponding to a second order or possibly smoother transition, described by 
a correlation length critical exponent $\nu_r\ge 2/d=2$ (with, to our 
knowledge, no theoretical 
prediction available for the $\nu_r$ value).

Whereas these theoretical determinations are appropriate to account for
the behaviour of the model in the thermodynamic limit, in the experimental 
situation we are typically concerned with the thermal denaturation of DNA 
molecules 
of specific sequence and given finite size. In such context it is implicitly 
assumed
that the intrinsic disorder is at the 
origin of the well-known multi-step behaviour 
of the density of closed base pairs, 
$\theta(\{\varepsilon_i\},N,T)$, the order parameter of the  
transition as accessible through optical absorbance measurements. 
It is then 
classical to analyze the derivative of this quantity with respect to 
temperature 
($d\theta(\{\varepsilon_i\},N,T)/dT$), which 
displays distinct peaks as reflecting the multi-step behaviour. 
At this level, from the statistical mechanics point of 
view, the first order character of the transition in the pure case is in 
agreement with the observation that the steps are typically very steep, hence 
the corresponding peaks very sharp. 
On the other hand, the fact that GC-rich regions tend 
to open at higher temperatures than AT-rich ones, suggests in a particularly 
evident way the possibility of {\em non self-averageness} for densities of 
extensive quantities, in the sense that the positions and the characteristics 
of the steps, and of the corresponding peaks, are strongly sequence-dependent. 

On general grounds, it is accordingly expected 
\cite{AhHa} that, for $N \rightarrow \infty$, 
the temperature range where such behaviour is observed shrinks around the 
critical temperature $T_c$, the transition being rounded. In fact, it is only 
at the critical point that the presence of a diverging correlation length 
breaks down the argument usually used to demonstrate self-averageness of 
densities of extensive quantities \cite{Br}, built on  
the description of the system as consisting of
weakly interacting sub-systems. 
However the validity of such 
picture is not obvious for the present almost one-dimensional polymer model, 
in which self-avoidance represents an infinite range interaction.

In the background of the recent 
mathematical findings \cite{GiTo1,GiTo2,Gi,To}, it is expected that
the effect of disorder 
on this peculiar first order transition would be the same than the one 
predicted 
theoretically for random ferromagnets \cite{AhHa} 
(as further confirmed numerically in \cite{WiDo} for site dilute Ising 
models), with in particular the mean value and the fluctuations of an 
appropriately defined pseudo-critical temperature scaling with the same 
exponent. Nevertheless it is highly desirable for the present model to further
assess numerically this picture, in order notably to better understand 
{\em the way in which} relevance of disorder becomes manifest, as tackled at 
the finite size level. 

\section{Model and observables}
\label{beginning}
\subsection{The pure case}
\label{beginning_one}
For the sake of completeness, we first recall briefly the salient features of 
the homogeneous on-lattice model which was introduced in \cite{CaCoGr}: two SAWs
with the same origin on a $3d$ cubic lattice, obeying the simple rule that two 
monomers can occupy the same lattice point (gaining a coupling energy 
$\varepsilon$) if and only if their positions along the two chains are 
identical. The thermal variables describing the system can be represented by 
the ensemble of the $\{s_i\}$, with $s_i=0$ if the base pair in position $i$ 
is in the open state and $s_i=1$ if it is in the closed state. The
possible configurations are only the ones allowed by self-avoidance, thus 
introducing an infinite range interaction of entropic nature in an almost 
uni-dimensional Ising model \cite{Fi0}. The energetic contribution of a given 
configuration to the Boltzmann weight of the ``spins'' $\{s_i\}$ (hence the 
associated Hamiltonian) is then simply written as 
${\cal H}=-\varepsilon \sum_{i=1}^{N}s_i$. It can be noticed that $\varepsilon$
enters in the partition function only through the ratio $\varepsilon/T$, 
allowing thus to set $\varepsilon=1$ without loss of generality.

This model can in particular be assimilated to a homogeneous DNA denaturation 
transition model {\em \`a la} PS \cite{PoSh1,PoSh2,Fi0,Fi1,CaCoGr}, in which 
the two chains represent the two strands of the double helix, allowing for 
one free-end but not base-pair mismatches. It is then interesting to 
notice that self-avoidance is fully taken into account in the on-lattice 
numerical simulations, by design. In fact, it is found \cite{CaCoGr} that,
in the thermodynamic limit of infinite chain length $N \rightarrow \infty$, 
the model undergoes a first order phase transition with a discontinuity in the 
order parameter, the density of closed base pairs $\theta(T)= 
\frac{1}{N} \sum_{i=1}^N \langle s_i \rangle$ ($\langle \cdot \rangle$ 
holding for the $thermal$ average), which varies abruptly from zero value at 
high temperature to a finite value below the critical temperature $T_c$. 
The properties of the system are better grasped in the behaviour of the 
loop-length probability distribution \cite{KaMuPe1,KaMuPe2,CaOrSt,BaCaSt,
BaCaKaMuOrSt}:
\begin{equation}
P(T,l) \propto \frac{e^{-l/\xi(T)}}{l^{c_p} }, 
\label{plh}
\end{equation}
in which appear both the exponent $c_p$, describing the purely algebraic decay 
of $P(T,l)$ at the critical point, and the correlation length $\xi(T)$.
In the considered case, in which we allow for one free-end, it could be 
necessary in 
principle to take into account an additional length 
scale, corresponding to the distance between end points \cite{CaCoGr,MoGa}. 
Nevertheless, such requirement should be basically considered to be associated 
with a boundary condition 
\cite{KaMuPe2}, therefore not influencing the thermodynamic limit 
properties of the model. 

In general, upon solving these models in the {\em grand canonical} ensemble, 
the succession of closed segments (helix) and open loops (coil regions) sums up 
as a geometric series \cite{Fi0}. However, in the present case, such treatment 
amounts to neglect the self-avoiding interactions between different segments 
and loops, which can be taken into account more correctly by the appropriate 
renormalization of $c_p$ \cite{KaMuPe1,KaMuPe2}. Transitions in homogeneous 
PS models are then found to be of first order, both in $d=3$ and in $d=2$. 
Accordingly, they are characterized by an order parameter critical exponent 
$\beta_p=0$, {\em i.e.} a discontinuity in $\theta(T)$, but also by the 
presence of a diverging correlation length, with 
related critical exponent $\nu_p=1$ (since renormalization leads to $c_p > 2$). 
In the theoretical framework of almost uni-dimensional phase transitions this 
correlation length is to be interpreted as the {\em longitudinal} one 
\cite{Fi0,KaMuPe2}, corresponding to the mean loop length 
$\langle l \rangle$ along the chain, whose variance diverges.
More in detail, based on conformal field theory results \cite{Du}, the 
theoretical prediction in $d=3$ for a self-avoiding loop embedded 
in a self-avoiding chain \cite{KaMuPe1,KaMuPe2} leads to a $c_p$ value between 
2.115 and 2.22, in very good agreement with the value $c_p \simeq 2.15$ found 
in numerical study of the on-lattice SAW DNA model 
\cite{CaOrSt,BaCaSt,BaCaKaMuOrSt}.

The behaviours at criticality expected for the singular parts of the various 
thermodynamic quantities in  PS models are summarized in [Tab. 1], defining 
implicitly the order parameter critical exponent $\beta_p$, the correlation 
length critical exponent $\nu_p$, and the specific heat critical exponent 
$\alpha_p$ (the label $p$ is to distinguish the pure $p$ case from the 
random $r$ one), all of them being expressible as functions of 
the sole variable $c_p$. It can be noticed in particular that here the energy 
behaves as the order parameter, hence the singular part of the specific heat 
behaves as the one of the derivative of the order parameter with respect to 
the temperature, as well as to that of its derivative with respect to 
$\varepsilon$, {\em i.e.} to the singular part of the susceptibility. 
Moreover $T_c$ is to be reached 
from the low temperature phase, as the thermodynamic limit correlation 
length is infinite in the whole phase $T>T_c$ (both with and 
without one free-end allowed).

\begin{table}
\begin{center}
$\begin{tabu}{|| l l  ||}
\hline
\hline
& \\
\lim_{T \rightarrow T_c^-}\theta(T) \propto (T_c-T)^{\beta_p} & 
\beta_p = \left \{
\begin{array}{c c c}
0 & \mbox{ for } & c_p \ge 2 \\
(2-c_p)/(c_p-1) & \mbox{ for } & c_p < 2 
\end{array}
\right. \\
& \\
\hline
& \\
\lim_{T \rightarrow T_c^-}\xi(T) \propto (T_c-T)^{-\nu_p} & \nu_p = \left \{
\begin{array}{c c c}
1 & \mbox{ for } & c_p \ge 2 \\
1/(c_p-1) & \mbox{ for } & c_p < 2 
\end{array}
\right. \\
& \\
\hline
& \\
\lim_{T \rightarrow T_c^-}\frac{\textstyle d\theta(T)}{\textstyle dT} \propto 
(T_c-T)^{-\alpha_p} & 
\alpha_p = 1-\beta_p=2-\nu_p \\
& \\
\hline
& \\
\lim_{T \rightarrow T_c^-}P(T,l) \propto 
\frac{1}{l^{c_p}} & \\
& \\
\hline
\hline
\end{tabu}$
\caption{Behaviours at criticality of the singular parts  of the relevant 
thermodynamic quantities in the model, which involve only one independent 
critical exponent (see text for details).}
\end{center}
\end{table} 

It can be useful to recall, in order to better specify the link between the 
on-lattice $3d$ SAW DNA model and the associated PS one, that the transition 
in the grand canonical ensemble occurs in fact at a tri-critical point 
in the fugacity-temperature plane (see \cite{CaCoGr} and references therein), 
better described by the crossover exponent $\phi_p=\nu_g^{3d}/\nu_p^{3d}$, 
where $\nu_g^{3d}=1/{\cal D}$ is the geometrical correlation length 
critical exponent (given by the inverse of the fractal dimension ${\cal D}$ of 
the $3d$ SAW), whereas $\nu_p^{3d}$ corresponds to the thermal 
correlation length critical 
exponent. Accordingly, the relation $\alpha_p=2-\nu_p$ (which in PS models 
corresponds simply to a particular case of the well known relation 
$\alpha_p=2-d\nu_p$ in $d=1$), is the analogous within the almost 
uni-dimensional framework to the $3d$ relation 
$\alpha_p=2-{\cal D}\nu_p^{3d}=2-\nu_p^{3d}/\nu_g^{3d}=2-1/\phi_p$ (with the 
first order case corresponding to $\phi_p=1$, {\em i.e.} 
$\nu_p^{3d}=\nu_g^{3d}$). On the other hand, the definition of $P(T,l)$ given by 
Eq.~(\ref{pl}) implicitly refers to the unidimensional structure of the 
system (with $l$ the loop length along the chain), and accordingly the 
corresponding correlation length critical exponent is given by 
$\nu_p={\cal D}\nu_p^{3d}=1/\phi_p$. It is worth noting that the fractal 
dimension ${\cal D}$ is by definition the same in the presence of disorder.

\subsection{The random case}
\label{beginning_two}
The DNA is intrinsically a disordered system, because of the difference 
in the coupling energies between $AT$ and $GC$ base pairs. To account for 
disorder it is then necessary to introduce a coupling dependent on the 
position $i$ of the base pair along a given sequence, $\varepsilon_i$, writing 
accordingly: 
\begin{equation}
{\cal H}=-\sum_{i=1}^{N}\varepsilon_i \: s_i.
\end{equation}
It is then important to distinguish 
{\em thermal} averages, $\langle ( \cdot ) \rangle$ (over the thermal 
variables $\{ s_i \}$), from {\em disorder} averages, $\overline{ (\cdot) }$ 
(over the different 
possible realizations of the quenched random variables $\{ \varepsilon_i \}$, 
{\em i.e.} the different possible sequences). 

A simple choice for the couplings \cite{Co,GaMo,MoGa,CoYe} consists in 
considering identically distributed independent random variables 
$\{ \varepsilon_i \}$, following the binomial law:
\begin{equation}
P(\varepsilon_i)=\frac{1}{2} \left [ \delta (\varepsilon_i - 
\varepsilon_{AT}) + \delta   (\varepsilon_i - 
\varepsilon_{GC}) \right ],
\label{disorderprobability}
\end{equation}
which assumes the same mean $AT$ and $GC$ contents,  
leaving free the choice of the value for the energy ratio parameter 
$R=\varepsilon_{GC}/\varepsilon_{AT}$.
In the present work, as in \cite{CoYe}, we study the PS model corresponding to 
the disordered on-lattice $3d$ SAW DNA in \cite{Co}, with $R=2$ (obtained by 
taking 
$\varepsilon_{GC}=2$ and $\varepsilon_{AT}=1$). 

In terms of recurrent evaluations, for a system of length $n+1$ with both base 
pairs $i=1$ and $i=n+1$ in the closed state ($s_1=s_{n+1}=1$), the 
{\em forward} partition function $Z^f(\{ \varepsilon_i \},n+1,T)$, which sums 
the contributions of all the configurations weighted by their Boltzmann 
factors, is obtained from $Z^f(\{ \varepsilon_i \},n,T)$ as:
\begin{equation}
Z^f(\{ \varepsilon_i \},n+1,T)=e^{(\beta \epsilon_{n+1}-\log\mu)} \left [
Z^f(\{ \varepsilon_i \},n,T)+ \sum_{n'=1}^{n-1} 
\frac{Z^f(\{ \varepsilon_i \},n',T)}{[2(n-n'+1)]^{c_p}} \right ].
\label{zf}
\end{equation}
Similarly, the {\em backward} partition function 
$Z^b(\{ \varepsilon_i \},n-1,T)$ is obtained as:
\begin{equation}
Z^b(\{ \varepsilon_i \},n-1,T)=e^{(\beta \epsilon_{n-1}-\log\mu)} \left [
Z^b(\{ \varepsilon_i \},n,T)+ \sum_{n'=n+1}^{N} 
\frac{Z^b(\{ \varepsilon_i \},n',T)}{[2(n'-n+1)]^{c_p}} + 1 \right ],
\label{zb}
\end{equation}
where the last term in the expression between parentheses is for taking into 
account the allowance for one free-end. 

The implementation of the model also implies a choice 
for the value of the connectivity constant $\mu$: 
as well known, $\mu=2d$ for a $d$-dimensional random 
walk on a hyper-cubic lattice and $\mu \simeq 4.7$ for a $3d$ SAW on a cubic 
lattice, leading to $\log \mu \simeq 1.54$ in our case \cite{CoYe}. 

\begin{table}
\begin{center}
$\begin{tabu}{|| r c  l r ||}
\hline
\hline
& & &\\
e(\{\varepsilon_i\},N,T)& \equiv - &\frac{\textstyle 1}{\textstyle N}  
\sum_{i=1}^N \varepsilon_i \langle s_i \rangle & \mbox{(energy density)}\\
& & &\\
\hline
& & & \\
c(\{\varepsilon_i\},N,T)& \equiv &\frac{\textstyle 1}{\textstyle T^2} 
\frac{\textstyle \partial}{\textstyle \partial T}e(\{\varepsilon_i\},N,T) & 
\mbox{(specific heat)} \\
& & & \\
\hline
& & & \\
\theta(\{\varepsilon_i\},N,T)& \equiv & \frac{\textstyle 1}{\textstyle N} 
\sum_{i=1}^N \langle s_i \rangle & \mbox{(order parameter)}\\
& & & \\
\hline
& & & \\
\chi(\{\varepsilon_i\},N,T)& \equiv & \left \{
\begin{array}{l}
\frac{\textstyle 1}{\textstyle N} \left [
\langle \left ( \sum_{i=1}^N s_i \right )^2 \rangle - 
\left ( \langle  \sum_{i=1}^N s_i \rangle \right )^2 \right ]  \\ 
T \left [ \frac{\textstyle \partial}{\textstyle \partial\epsilon_{AT}} 
\theta(\{\varepsilon_i\},N,T) +  
\frac{\textstyle \partial}{\textstyle \partial\epsilon_{GC}} 
\theta(\{\varepsilon_i\},N,T) \right ]  \\  
\end{array} \right.
&  \mbox{(susceptibility)} \\
& & & \\
\hline
\hline
\end{tabu}$
\label{observables}
\caption{Definition of relevant thermodynamic observables in the
model.}
\end{center}
\end{table} 

The total partition function for a sequence $\{ \epsilon_i\}$ of length $N$ is 
given by: 
\begin{equation}
Z(\{ \varepsilon_i \},N,T)=\sum_{n=1}^{N}
Z^f(\{ \varepsilon_i \},n,T)=Z^b(\{ \varepsilon_i \},1,T). 
\end{equation}
Therefore, it is possible to evaluate the probability for a base pair
in position $n$ to be in the closed state ({\em i.e.} the thermal average 
$\langle s_i \rangle$): 
\begin{equation}
{\cal P}(\{ \varepsilon_i \},N,T,n)= \langle s_i \rangle= 
\frac{Z^f(\{ \varepsilon_i \},n,T)Z^b(\{ \varepsilon_i \},n,T)}
{Z(\{ \varepsilon_i \},N,T) \exp({\beta \epsilon_n-\log\mu})},
\label{Ps}
\end{equation}
from which relevant thermodynamic observables (whose definitions are recalled 
in [Tab. 2]) are obtained easily. It is worth 
noticing that results in \cite{Co,CoYe} 
are in agreement with a situation in which, also in the presence 
of disorder, the order parameter behaves as the energy 
(with $\beta_r=\nu_r-1$), 
and the susceptibility as the specific heat (with $\alpha_r=2-\nu_r$). 
Under such 
conditions only one independent critical exponent needs to be evaluated. 

Finally, we recall that the loop-length probability distribution, 
apart from the normalization constant (imposing 
$\sum_l P(\{ \varepsilon_i \},T,N,l)=1$),
can be evaluated as:  
\begin{equation}
P(\{ \varepsilon_i \},N,T,l) \propto \frac{1}{l^{c_p}} 
\sum_{n}\frac{Z^f(\{ \varepsilon_i \},n,T)Z^b(\{ \varepsilon_i \},n+l+1,T)}
{Z(\{ \varepsilon_i \},N,T)}.
\label{pl}
\end{equation}
This expression further highlights the fact that in the present model $c_p$ 
corresponds to an $input$.

In this context, the simplest picture describing the random fixed point, 
in agreement with the numerical results in \cite{Co,CoYe}, is 
the one in which in the thermodynamic limit:
\begin{equation}
P(T,l)=\lim_{N \rightarrow \infty} \overline{
P(\{ \varepsilon_i \},N,T,l)} \propto \frac{e^{-l/\xi(T)}}{l^{c_r}},
\label{plscaling}
\end{equation}
Accordingly, as in the pure case (see Eq.~(\ref{pl})), the average loop-length 
probability distribution is expected to display a purely algebraic decay at the 
critical point, where the average correlation length $\xi(T)$ diverges. This
purely algebraic decay is described by the random critical exponent $c_r$, 
which in the case of a smooth transition should be linked to the correlation 
length critical exponent $\nu_r$ by the relation $c_r=1+1/\nu_r$: the same kind 
of relation which is known to be valid for the pure system, for $c_p<2$. 

\subsubsection{Implementation of the PS model with self-avoidance in various 
numerical studies}
\label{beginning_twob}
The PS model considered here corresponds in a 
detailed way to the on-lattice $3d$ disordered SAW DNA model, with 
the same distribution of the coupling energies, studied numerically in 
\cite{Co}. 
In this work, by means of parallel 
computing, it was possible to collect sufficient statistics up to chain 
lengths $N=800$. 
Simulations were extensively performed in the case $R=2$, considering also 
different 
values of $R$. With such choices of the parameters,  
applying standard finite size scaling, it was found that the length scale 
considered was not large enough for reaching conclusive results. 

More in detail,
the estimations obtained for $\phi_r$, which appeared 
moreover to depend on $R$, were 
smaller than $\phi_p=1$, and correspondingly the ones 
for $c_r$ were smaller than $c_p \simeq 2.15$ (more
precisely smaller than 2). 
Nevertheless, both estimations for the exponents 
were still compatible with the pure case values, within the errors. In 
addition, no conclusive results were obtained 
when fitting data on the loop-length probability
distribution at different temperatures, nor when attempting to make an 
analysis in terms of pseudo-critical temperatures.  
In any event, these estimations clearly suggested a 
value for the 
correlation length critical exponent 
definitely smaller than the 
value $\nu_r= 2$,  {\em i.e.} the smallest possible value 
which would be expected in the case of relevance of disorder on general grounds 
\cite{Ha,ChChFiSp,AiWe,Be}  
with such conclusion further confirmed 
for a class of polymer models 
encompassing the present one \cite{GiTo1,GiTo2,Gi}.

\begin{table}
\begin{center}
$\begin{tabu}{|| c | c | c | c ||}
\hline
\hline
&  \begin{array}{l}\mbox{on-lattice} \\
\mbox{SAW model} \\ \mbox{in \cite{Co}} \end{array}  & 
\begin{array}{l}\mbox{PS model} \\
\mbox{with $c_p=2.15$} \\ \mbox{in \cite{GaMo,MoGa}} \end{array} &
\begin{array}{l}\mbox{PS model} \\
\mbox{with $c_p=2.15$} \\ \mbox{here and in \cite{CoYe}} \end{array}
\\
\hline
\hline
\mbox{energy ratio } R & 2 &  1.098 & 2 \\ 
\hline
\mbox{connectivity constant } \mu & \simeq 4.7 & 2 & 4.7 \\
\hline
\begin{array}{l}\mbox{cooperativity factor} \\
\mbox{(for the loops)} \end{array} \sigma_0 & 1 & 0.296 & 1 \\
\hline
\begin{array}{l}\mbox{cooperativity factor} \\
\mbox{(for the free end)} \end{array}\sigma_1 & 1 & 0.5 & 1 \\
\hline
\mbox{free-end exponent } c'_p & < 10^{-1} & 0 & 0 \\ 
\hline
\hline
\end{tabu}$
\label{parameters}
\caption{Values of parameters in the 
various numerical studies of disordered PS models for DNA 
denaturation taking into account self-avoidance. The cooperativity factor 
$\sigma_0$ is to account for the cost of opening a loop; it was predicted in 
\cite{KaMuPe2} that the free-end exponent $c'_p$ takes a small value.}
\end{center}
\end{table} 

This overall situation appeared then all the more unsettled with the
numerical results reported in \cite{GaMo,MoGa} for a PS model with 
$c_p=2.15$ supporting the alternative scenario
of a pseudo first order transition. 
In the perspective of a global clarification of the situation, taking
into account our results here and in the previous work \cite{CoYe}, 
[Tab. 3] reports the values of the parameters adopted in 
the various studies. This table highlights notably the differences in the 
values 
of the energy ratios $R$ and of the connectivity constants $\mu$, 
the importance 
and significance of such differences being further discussed below.

\subsection{Scaling laws and pseudo-critical temperatures}
\label{beginning_three}
In the presence of a phase transition characterized by a diverging correlation 
length it is well known that the critical exponents can be evaluated 
numerically 
using finite size scaling techniques \cite{Ba,SeShTa,BiYo}. Such evaluations 
are 
based on the theoretical argument, that can be shown to be valid in the 
framework 
of the renormalization group approach \cite{ZJ}, following which the only 
relevant adimensional ratio near the critical point is the one between the 
thermodynamic limit correlation length itself, $\xi(T) \sim |T_c-T|^{-\nu}$, and 
the linear scale $L$ of the system under consideration (namely in the present 
$1d$ case $N=L$). In a system without quenched disorder ($\nu=\nu_p$) this 
means 
that an observable ${\cal O}$, with thermodynamic limit behaviour of its 
singular 
part described by the critical exponent $e_p$ 
($\lim_{T \rightarrow T_c}{\cal O}(T) \propto |T_c-T|^{e_p}$), is expected to 
follow 
the law:
\begin{equation}
{\cal O}(L,T)=L^{-e_p/\nu_p} \tilde{{\cal O}}[(T_c-T)L^{1/\nu_p}],
\label{scalinga}
\end{equation}
with $\tilde{{\cal O}}$ a scaling function.

As discussed for instance in \cite{SeShTa}, in the case of the diluted 
Ising model (a quenched disordered system which is similar to the present one,
notably relative to the fact that it does not display 
competing interactions, and accordingly no frustration),  
an analogous scaling picture is expected to describe random critical 
points, and should therefore
allow to evaluate the correlation length critical 
exponent $\nu_r$, as well as the critical exponent $e_r$ 
related to the considered observable. 
Such evaluation can be performed by 
studying the scaling behaviour 
of $\overline{{\cal O}(\{ \varepsilon_i \},N,T)}$, 
with the average over the quenched variables 
$\{ \varepsilon_i \}$ estimated in the standard way, 
taking the same temperature for all ${\cal N}_s$ samples (here sequence 
realizations):
\begin{equation}
\overline{{\cal O}(\{ \varepsilon_i \},N,T)}= \frac{1}{{\cal N}_s}
\sum_{\{ \varepsilon_i \}} {\cal O}(\{ \varepsilon_i \},N,T)
\label{standardaverages}
\end{equation}
Such standard finite size scaling technique was 
applied 
both for the study of the on-lattice disordered SAW DNA model \cite{Co} and for 
the corresponding PS model with $c_p=2.15$ \cite{CoYe}. In this last work, 
involving large enough statistics with chain lengths up to 
$N=2\cdot10^4$, such 
approach provided notably numerical evidence for relevance of disorder, with 
the random critical point of the system described by a correlation length 
exponent $\nu_r = 2.9 \pm 0.6$. 

However, it was also generally put forward 
\cite{WiDo} that, when considering random critical 
points, results of standard finite size scaling analyses should be considered 
with some care. 
Within this framework, following a general renormalization 
group result for random ferromagnets \cite{AhHa}, we are led to introduce an 
additional 
observable, the effective critical temperature (or pseudo-critical 
temperature), $T_c(\{ \varepsilon_i \},N)$, studying its dependence on the 
disordered configurations considered. In the present case ($d=1$ ) with $L=N$, 
the mean value of this quantity:
\begin{equation}
 T_c(N) \equiv \overline{T_c(\{ \varepsilon_i \},N)}, 
\end{equation}
and its fluctuations:
\begin{equation}
\delta T_c(N) \equiv 
\left \{
\overline{\left[T_c(\{ \varepsilon_i \},N)\right]^2}-
\left[\overline{T_c(\{ \varepsilon_i \},N)}\right]^2 \right \}^{1/2},
\end{equation}
are expected to behave as functions of the system size \cite{AhHa,WiDo}:
\begin{eqnarray}
\left.
\begin{array}{lcl} 
T_c(N) & \simeq & T_c + C N^{-1/\nu_p} \\
\delta T_c(N) & \propto & N^{-1/2} \\
\end{array} \right \} 
\mbox{ for irrelevant disorder } 
\label{pseudotcirrev}
\\
\left. 
\begin{array}{lcl} 
T_c(N) & \simeq & T_c + C N^{-1/\nu_r} \\
\delta T_c(N) & \propto & N^{-1/\nu_r} \\
\end{array}
\right \} \hspace{.1in}
\mbox{ for relevant disorder   } 
\label{pseudotcrev}
\end{eqnarray} 
Noticeably, the theoretical framework \cite{AhHa} predicts that, for 
relevant disorder, the mean value and the fluctuations 
of the pseudo-$T_c$ should 
scale with the same exponent. This same framework
also allows to infer that disorder should be relevant 
as soon as the specific heat critical 
exponent for the pure system fulfills the condition $\alpha_p>0$ 
(hence, from the hyper-scaling
relation $\alpha_p=2-d \: \nu_p$, as soon as the correlation length 
critical exponent of the pure system fulfills $\nu_p < 2/d$). 

On the other hand, 
it is worth noting that the behaviour of the mean value and 
of the fluctuations of the pseudo-$T_c$ described by two different exponents in
Eq.~(\ref{pseudotcirrev}), corresponding to a situation 
encountered usually when disorder is irrelevant (as the specific heat of the 
pure system displays no divergence for $\alpha_p \le 0$), could 
be also attributed in fact to the presence 
of two correlation lengths. 
The theoretical basis
for such a possibility was laid within the renormalization
group framework, notably in the case of random transverse field Ising chains 
\cite{DFi}.
Furthermore, in addition to the results on the PS model with self-avoidance in 
\cite{GaMo,MoGa}, evidence for such scenario, 
corresponding to a pseudo first order transition, was reported in various 
other cases 
(see for instance \cite{IgLiRiMo} and \cite{MoGa2}). 

In such context, in the case of the PS model with self-avoidance, 
an independently diverging correlation length could be 
pictured as related to the free-end distance,
or, perhaps on more grounded bases, it could be 
hypothesized that the divergence of the typical loop is different from that 
of the average one. Accordingly, the transition would be of first order, as in 
the pure case, from the point of view of the behaviour of the {\em 
typical} observables (the given sequence undergoes a 
first order transition, with $\nu_{r,1}=1$) and it would be of second
order from the point of view of the {\em average} ones (whose behaviour
is ruled by $\nu_{r,2}=2$) \cite{GaMo,MoGa}. 

In fact, it is also 
obvious within this 
framework \cite{WiDo,GaMo,MoGa}, that the standard scaling law describing the 
finite size behaviour of a thermodynamic observable ${\cal O}$ with critical 
exponent $e_r$ 
is expected to 
be better obeyed by the quantity 
$\overline{\overline{{\cal O}}}$:
\begin{equation}
\overline{\overline{{\cal O}(\{ \varepsilon_i \},N,T)}}=
N^{-e_r/\nu_r}\tilde{{\cal O}} \left [ ( T_c(N)-T) N^{1/\nu_r} \right ],
\label{scaling}
\end{equation}
in which we label by $\overline{\overline{(\cdot)}}$ the average over 
disorder performed by taking into account the 
sequence-dependent $T_c(\{ \varepsilon_i \},N)$, according to: 
\begin{equation}
\overline{\overline{{\cal O}(\{ \varepsilon_i \},N,T)}}= 
\frac{1}{{\cal N}_s} \sum_{\{ \varepsilon_i \}} {\cal O} 
\left [ \{ \varepsilon_i \},N,T-T_c(\{ \varepsilon_i \},N)+T_c(N) \right ] 
\label{newaverages}
\end{equation}
In such a way, one avoids that the results 
are governed by the fluctuations of the pseudo-$T_c$. 

Finally, in the case of disorder relevance, theoretical results \cite{AhHa} 
imply {\em strong} non self-averageness in the thermodynamic observables 
which are {\em singular} at the critical point, despite of these being 
densities of extensive quantities.  
By definition, self-averageness is measured from the ratio: 
\begin{equation}
{\cal R}_{\cal O} = \frac{\overline{\left [ {\cal O}
(\{ \varepsilon_i \},N,T) \right ]^2}-
\left  [ \overline{{\cal O}(\{ \varepsilon_i \},N,T)} \right]^2}
{\left  [ \overline{{\cal O}(\{ \varepsilon_i \},N,T)} \right ]^2},
\label{nsa}
\end{equation}
with the 
strong non self-averaging behaviour corresponding to ${\cal R}_{\cal O} \sim 1$ 
(whereas ${\cal R}_{\cal O}\sim 1/N$ in the usual self-averaging behaviour). 
Noticeably,
these results for observables such as the order parameter
and the susceptibility are expected in the present model
both in the case where the mean value and the fluctuations of the 
pseudo-critical 
temperature scale with the same exponent and in the case of a pseudo first 
order transition.

Indeed, this behaviour has been observed in 
the order parameter in the PS models with different $c_p$ values studied in 
\cite{MoGa}, and in particular in the peculiar case of $c_p = 2.15$.
On the other hand, for the PS model here (as already considered in 
\cite{CoYe}) as
well as for the on-lattice SAW DNA (as considered in \cite{Co}),
such behaviour was clearly suggested by the presence of
multiple steps in the order parameter, and correspondingly of several peaks 
in the 
susceptibility, in a not negligible fraction
of the sequences,
already for relatively small chain lengths.

Within this background, a rather subtle 
point concerns the definition of the pseudo-$T_c$ itself.  
From this point of view, in \cite{GaMo,MoGa} 
two different definitions were 
introduced, with their study leading 
to the same conclusions: one is obtained from 
the free energy behaviour and the other, which we also study, as the 
crossing point of the order parameter $\theta(\{ \varepsilon_i \},T,N)$ 
curves, considering sequences of 
increasing lengths, obtained by the concatenation of 
increasing number of copies of a given original sequence. 

Importantly, the absence of multi-step behaviour in the order parameter in 
\cite{GaMo,MoGa}, for the PS model with $c_p=2.15$, represents the main 
qualitative difference with the results obtained for the on-lattice disordered 
SAW DNA or for the corresponding PS model in \cite{Co,CoYe}. 
Indeed, to the best of our knowledge, from this point of view the present work 
should represent the first attempt to define a pseudo-critical temperature 
with the order parameter displaying a multi-step behaviour. Noticeably, 
various definitions of the pseudo-$T_c$ have been considered, such as for 
spin glasses (\cite{Bietal}) 

On the other hand, in \cite{WiDo} the pseudo-$T_c$ was defined as 
the temperature corresponding to the {\em unique} maximum of the 
susceptibility. We are led to generalize this definition, and to take as 
pseudo-$T_c$ the temperature at which the susceptibility reaches its 
{\em absolute} maximum (also checking the agreement of the scaling
law with the alternative definition involving the concatenated sequence 
procedure). This choice, made possible by the accurately measured temperature 
dependence of the observables in the present study, appears 
generally to be the most reasonable one
and it should be moreover particularly appropriate to allow the 
identification of a possible 
pseudo first order transition. 
In fact, when defining the pseudo-$T_c$ in this way, by applying 
Eq.~(\ref{newaverages}) one finds:
\begin{equation}
\max_T \{{\overline{\overline{\chi(\{ \varepsilon_i \},N,T)}}} \}=
\overline{\max_T \{ \chi(\{ \varepsilon_i \},N,T) \}}.
\label{chimax}
\end{equation}
Accordingly, such definition should be 
effective for providing evidence for a diverging behaviour of the 
typical susceptibility (hence, in the present case in 
which the susceptibility behaves as the specific heat, 
for a specific heat exponent $\alpha_{r,1} = 1 = \alpha_p$,  
and also, by hyper-scaling in $d=1$, for $\nu_{r,1} = 1 = 
\nu_p$).

It was further hypothesized in \cite{MoGa} that the presence of two 
correlation lengths ruled by different critical exponents could be inferred
from the probability distribution of the loop lengths. Qualitatively, 
one would expect in particular different behaviours for 
$\overline{\log P(\{ \varepsilon_i \}, N,T)}$ and 
$\log \overline{P(\{ \varepsilon_i \}, N,T)}$, respectively. Even though such 
conclusion did not appear to be confirmed from the results in \cite{CoYe}, 
with the data at $T=T_c$ not displaying asymptotically detectable differences, 
it appears desirable to further investigate this point in detail, with careful 
analysis of data on the whole temperature range. 

\subsection{The SIMEX scheme and numerical implementations}
\label{beginning_four}
For efficient numerical implementation, the recursive equations for the 
forward and backward partition functions in the PS model are solved with the 
SIMEX scheme (SIMulations with EXponentials), resorting to an approximation of 
the power law $1/l^{c_p}$ with a sum of exponentials. The basic idea in the 
SIMEX scheme, as used in the context here \cite{CoYe,GaMo,MoGa}, was
originally expressed in  \cite{FiFr}, 
specifically for the numerical study of PS models of denaturation transitions
in linear DNA molecules. However, the generality of the powerful idea 
at the basis of this 
representation of the long-range effect 
as a sum of exponentials was not appreciated to its fair value, as in 
the original work \cite{FiFr} it was implemented in the context of conditional 
probabilities specific to the model considered.

The formulation of the SIMEX proper proceeded then in two steps. First the 
original idea was reformulated in more general terms for the linear PS model, 
in the context of recursions written 
directly in terms of partition functions \cite{Yeetal}. It was then possible, 
on such basis, to propose generalizations of the idea to higher order models 
involving several mutually coupled long-range effects \cite{Ye1}, with the 
corresponding algorithmic complexities reduced by several orders of magnitude. 
In order to be effective such a method relies on the accurate representation 
of long-range effects (such as the probability law for the lengths 
of the loops, in the simplest case of the model here) as sums of 
exponentials. The Pad\'e-Laplace method \cite{YeCl,ClDeYe} provides an elegant 
generic solution to this problem, not only in the case of  
purely decaying functions, such as power-laws (represented as sums of real 
exponentials), but more 
generally allowing to represent functions with complex 
exponentials. As an illustration for potential applications, the SIMEX 
scheme was used for example to implement sequence 
alignments with realistic gap models, in bioinformatics 
\cite{YeDe}.

In this background, with the approximation here,
\begin{equation}
\frac{1}{(2l)^{c_p}} \simeq \sum_{k=1}^{{\cal N}_E}a_k \: e^{-2lb_k},
\label{exprap}
\end{equation}
it is important to stress that, from the analytic point of view, replacing the 
power-law with exponentials is expected to change the nature 
of the singularity in the thermodynamic limit partition function. However, on 
numerical grounds, such replacement is 
not expected to influence the results of finite size scaling analyses.
Indeed, with the accuracy of the multi-exponential representation adopted (see 
below), the numerical approximation is practically indistinguishable 
({\em i.e.} within prescribed reasonable limits) from the 
analytic expression, well beyond the largest length scale considered. 

In this direction, because of the importance of the underlying
numerical problem, it is relevant to further recast the obtention of 
multi-exponential representations in the general context of approximation 
problems.
Since Prony (1795) the problem of obtaining numerical representations  of 
given functions as sums of exponentials has been addressed in many  fields and 
contexts. As a matter of fact, this problem can be declined  in two very 
different -in principle- flavours: identification or approximation. In the 
identification case the given function is  supposed, by essence, to be a sum 
of exponentials and the problem  consists in retrieving precisely the 
{\em genuine} number of exponential  components, with the estimates of the 
associated parameters. This  problem is reputed difficult, notably in the case 
of real  exponentials, because of ill-conditioning. It is then easy to 
{\em over-fit} the data, with methods such as least-squares, with increasing  
number of components, thus missing the aim of proper identification of  
the model in terms of its original components. In the alternative  case, 
related to approximation, the problem is of course completely  different, 
as the given function, known analytically, is not a sum of  exponentials and 
the aim is to obtain indeed the best possible  approximation under such form.
The present situation, concerning the power-law for the long-range  effect, is 
of course relevant to the approximation case, and we need  to ensure an 
accurate representation of the power-law up to the  largest considered sizes 
for the system with the sum of exponentials.  The number of components in the 
multi-exponential representation will then of course depend on the maximal 
size of the system considered,  with the need to introduce, according to
this size, 
additional components with increasingly smaller $b_k$ parameters  
(see Eq.~(\ref{exprap})): in the numerical fit, for increasingly larger values  
of the variable $l$, close to the maximal one in the model, the  exponential 
components with the smallest $b_k$ parameters are required to decay according 
to the corresponding values of the power-law.

In such context, the Pad\'e-Laplace method (which encompasses various  other 
formulations such as the Prony method or the method of moments  as particular 
cases) was originally formulated for the identification  problem. It was 
however also used in the approximation context, for  the obtention of 
numerical approximations of the power-law with sums  of exponentials (see with 
this respect [Fig. 4] in reference \cite{YeCl}).  Interestingly, in such case, 
an {\em identification-like} behaviour was  observed, concerning the number of 
exponential components in the  representation: more precisely, it was observed 
that the number of  components appropriate for systems of size $N_{max}$ 
followed essentially a law in $\ln(N_{max})$ (10 components for 
$N_{max}$ up to $2 \cdot 10^4$, 14-15  components for $N_{max}$ up to $10^6$, 
such as 
in the genomic analyses \cite{Ye,YeJo,YeBoLa}). Of  course, within the 
approximation context, the 
original model for  the  multi-exponential representation as obtained by the 
identification  procedure can be further used as {\em initial guess} for 
least-square  fits, for further refined approximation. In various studies, 
including  in the original Fixman-Freire paper \cite{FiFr}, multi-exponential 
representations  for the power-law were obtained resorting to different 
methods. It is however interesting to notice that increasingly more  
complex models implemented in the context of the generalizations of  
the SIMEX, could involve non purely decaying general long-range  effects. 
In such case it would then be necessary to resort to  approximations with 
sums of general complex exponentials, as allowed  by the Pad\'e-Laplace 
method \cite{YeCl,ClDeYe}.

\begin{table}
\begin{center}
$\begin{tabu}{||c | c | c | c||}
\hline
\hline
N & {\cal N}_S & T_{min} & T_{max} \\
\hline
\hline
100 & 2000 & 0.95 & 1.2 \\
\hline
200 & 2000 & 0.95 & 1.2 \\
\hline
500 & 2000 & 1.0 & 1.16 \\
\hline
750 & 1000 & 1.0 & 1.16 \\
\hline
1000 & 1000 & 1.0 & 1.15 \\
\hline
2500 & 1000 & 1.02 & 1.14 \\
\hline
5000 & 1000 & 1.02 & 1.14 \\
\hline
7500 & 1000 & 1.04 & 1.12 \\
\hline
10000 & 600 & 1.04 & 1.14 \\
\hline
15000 & 500 & 1.04 & 1.12 \\
\hline
20000 & 500 & 1.04 & 1.12 \\
\hline
\hline 
\end{tabu}$
\caption{Number of sequences (${\cal N}_S$) and temperature intervals 
($[T_{min},T_{max}]$) adopted in the computations, following
the chain lengths ($N$).}
\label{TNs}
\end{center}
\end{table}

Here, in order to reproduce accurately the behaviour of 
$1/(2l)^{2.15}$, we implement a 
SIMEX scheme with ${\cal N}_E=15$ exponentials \cite{CoYe}, adopting the same 
representation (in term of values for the coefficients 
$\{(a_k,b_k) \: k=1,\dots,{\cal N}_E \}$)
than the one in \cite{GaMo,MoGa}, which proved to be adequate for chain 
lengths of order 
$N = 2 \cdot 10^6$. Further details for the numerical 
implementation of the SIMEX scheme are provided in the Appendix in \cite{CoYe}. 

Summarizing, we study  
the disordered PS model with $c_p=2.15$ by adopting
the values $R=2$, $\sigma=1$ and $\log \mu = 1.54$. 
The conditions chosen to collect statistics for each chain 
length $N$ are recalled in [Tab. 4] \cite{CoYe}, in terms
of number ${\cal N}_S$ of different sequences $\{ \varepsilon_i \}$ used and 
range of temperature intervals considered ($[T_{min},T_{max}]$; the temperature 
intervals being always divided into ${\cal N}_T=250$ equally spaced 
sub-intervals). 

For each sequence, the specific heat is evaluated 
by computing numerically the derivative of the energy 
density with $\delta T=(T_{max}-T_{min})/{\cal N}_T$, and for the numerical 
computations of the derivatives in the susceptibility the 
values $\delta \varepsilon_{AT}= \beta \: 10^{-4}$ and 
$\delta \varepsilon_{GC}=R \: \delta \varepsilon_{AT}$ are adopted. 
It was checked that such choices ensure appropriate accuracy for the 
calculations (with notably the errors on the positions of maxima 
consistently smaller than the sample-to-sample fluctuations) \cite{CoYe}. 
Finally, the errors on average quantities are evaluated from sample-to-sample 
fluctuations. 

\section{Study with pseudo-critical temperatures: results and insights}
\label{discussion1}
\subsection{Pseudo-$T_c$: definitions and properties}
\label{discussion1_one}
\begin{figure}[hpbt]
\begin{center}
\leavevmode
\epsfig{figure=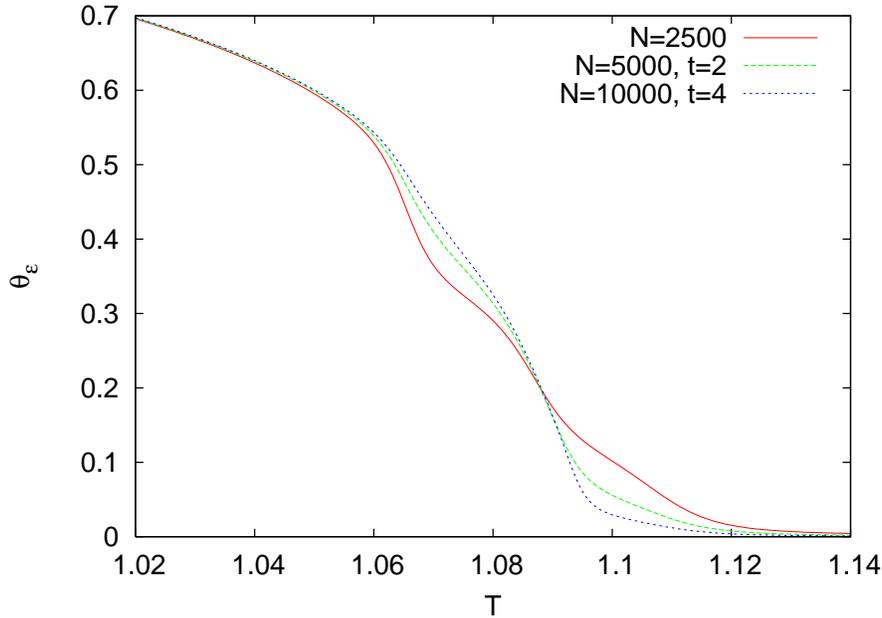,angle=270,width=12cm}
\caption{Order parameter $\theta$ plotted as function of the temperature for 
a given sequence of length $N=2500$ and for the sequences of lengths $N=5000$ 
and $N=10000$ obtained by concatenating $t=2$, respectively $t=4$, copies of 
the original sequence.}
\label{rev_new_fig1}
\end{center}
\end{figure}

\begin{figure}[hpbt]
\begin{center}
\leavevmode
\epsfig{figure=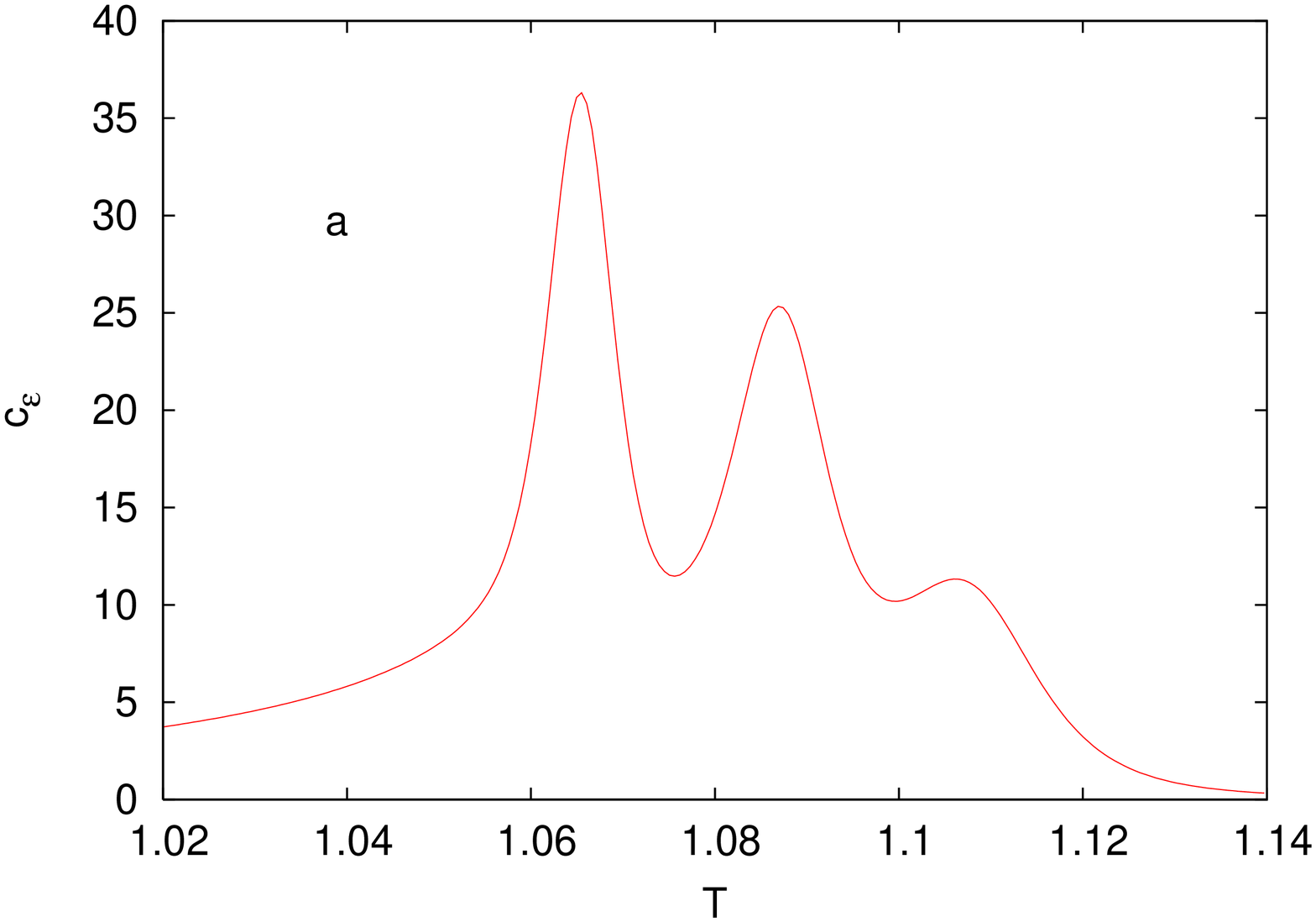,angle=270,width=8cm}
\epsfig{figure=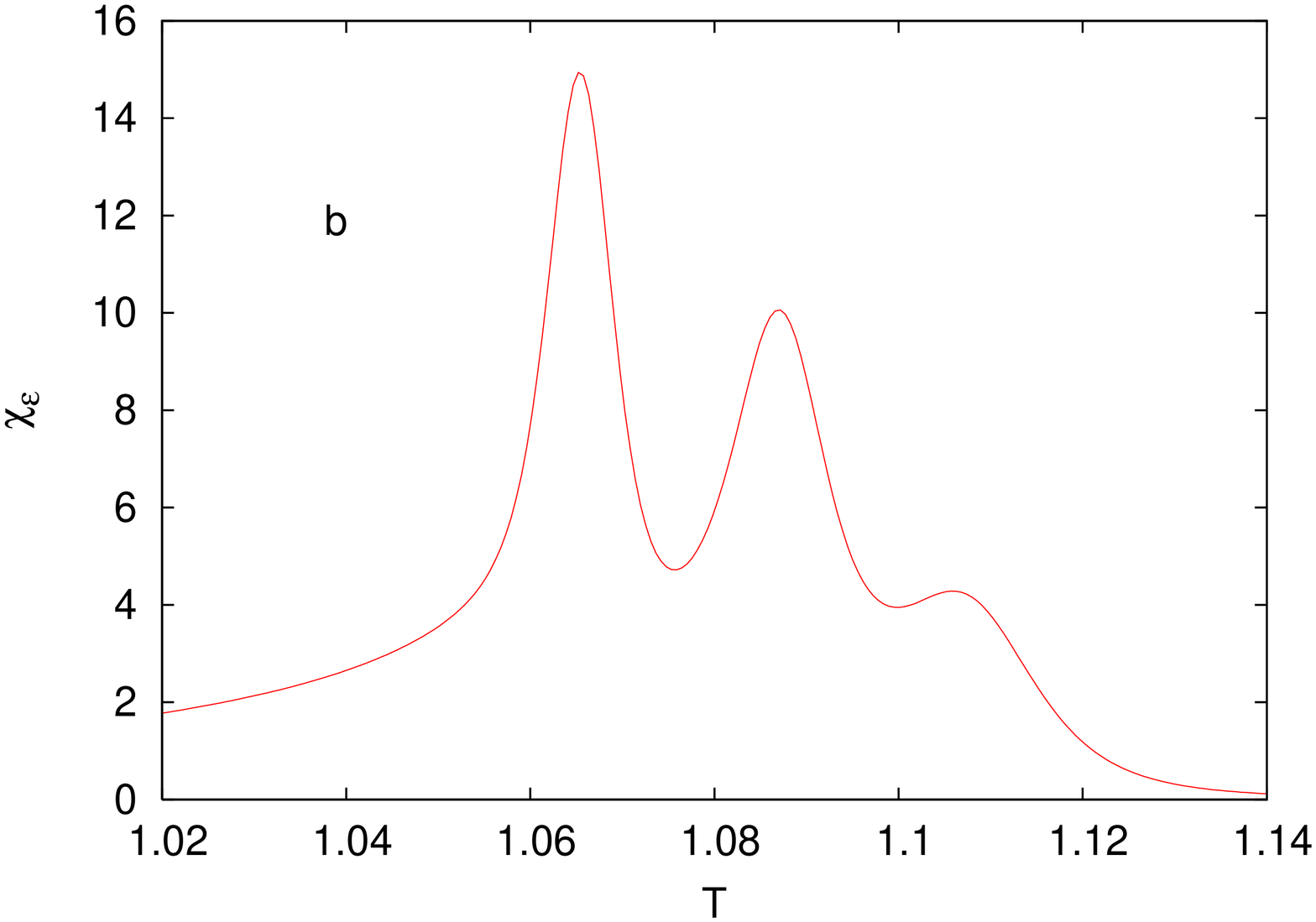,angle=270,width=8cm}
\caption{a) Specific heat $c$; b) susceptibility 
$\chi$. Data are for the same given sequence of length $N=2500$ considered
in the previous [Fig. \ref{rev_new_fig1}], plotted as 
function of temperature.}
\label{rev_new_fig2}
\end{center}
\end{figure}

Plot of the order parameter $\theta$ as function of the temperature 
is represented in [Fig. \ref{rev_new_fig1}], 
for a given sequence of length $N=2500$, and for 
sequences obtained by concatenating $t=2$ and 
$t=4$ times the original sequence. In the following [Fig. \ref{rev_new_fig2}],
we present data on the specific heat $c$ and the susceptibility $\chi$
for the same given sequence of length $N=2500$.
We can observe in [Fig. \ref{rev_new_fig1}] a clearly defined multi-step 
behaviour in the order parameter plot, with correspondingly three distinct 
peaks in the specific heat and in the susceptibility plots
in [Fig. \ref{rev_new_fig2}], even for this 
relatively short sequence considered.

We first notice, comparing the behaviours in [Fig. \ref{rev_new_fig2}a]
and [Fig. \ref{rev_new_fig2}b], that the pseudo-critical temperature 
$T_c(\{ \varepsilon_i \},N)$, defined as the temperature at which the 
susceptibility reaches its absolute maximum, is very close to the 
corresponding one obtained from the specific heat plot. This conclusion holds 
independently of the choice of a specific sequence. In fact, as already 
observed in \cite{CoYe}, for the parameter values used 
(notably ratio $R=2$, between energies associated with $GC$ and $AT$ links), 
for each 
disordered configuration, the specific heat, the derivative with respect to 
temperature of the order parameter 
and the susceptibility always display very similar behaviours. 
In particular, as detailed below, it appears that the disorder
averages of specific heat and  
of susceptibility (referring either to 
standard averages $\overline{(\cdot)}$ or to averages 
which take into account the sequence-dependent 
pseudo-critical temperature $\overline{\overline{(\cdot)}}$)  
display the same critical behaviour at $T_c$ and the same 
kind of finite size corrections to scaling. 
  
Interestingly, despite the multi-step 
behaviour, we observe in [Fig. \ref{rev_new_fig1}] that the densities of 
closed base pairs  
$\theta(\{ \varepsilon_i \},N,T)$, $\theta(\{ \varepsilon_i \},2N,T)$ and 
$\theta(\{ \varepsilon_i \},4N,T)$, respectively, cross at a well defined 
$T$-value, thus providing, following \cite{GaMo,MoGa}, a different possibility 
for defining a 
pseudo-critical temperature $T_{c}'(\{ \varepsilon_i \},N)$, which we also 
study here. It appears reasonable 
to assume that the scaling properties of this observable should not depend on 
the particular way it is defined. However, 
to the best of our knowledge, 
there are no previous attempts  
for defining and studying in detail pseudo-critical temperatures
in the presence of several distinct peaks in the specific heat 
or in the susceptibility and our definition
of $T_c(\{ \varepsilon_i \},N)$ as the position of the absolute maximum
of the susceptibility appears reasonable in such case. 
In any event, in the present 
work, we checked that with our choice
we get results compatible with those obtained with the 
alternative definition above, for chain lengths up to $N=2500$. 

\begin{figure}[hptb]
\begin{center}
\leavevmode
\epsfig{figure=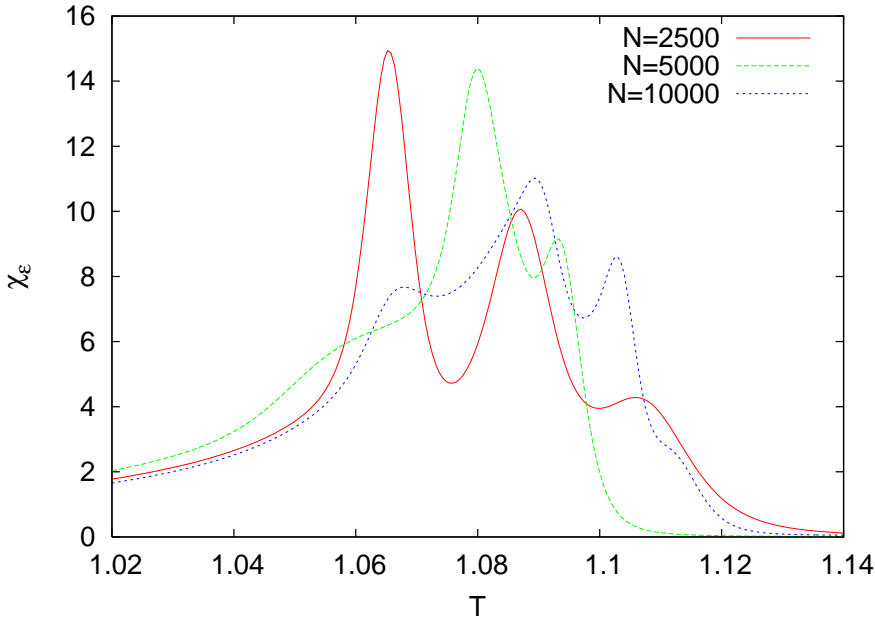,angle=270,width=12cm}
\caption{Susceptibility as function of temperature, plotted for 
independently randomly generated sequences
of lengths $N=2500$, $N=5000$ and $N=10000$, respectively. The randomly 
generated sequence of length $N=2500$ is the same than the one used in 
the previous figures, hence the corresponding susceptibility is in particular 
the same than in [Fig. \ref{rev_new_fig2}b].}
\label{rev_new_fig3}
\end{center}
\end{figure}

\begin{figure}[hptb]
\begin{center}
\leavevmode
\epsfig{figure=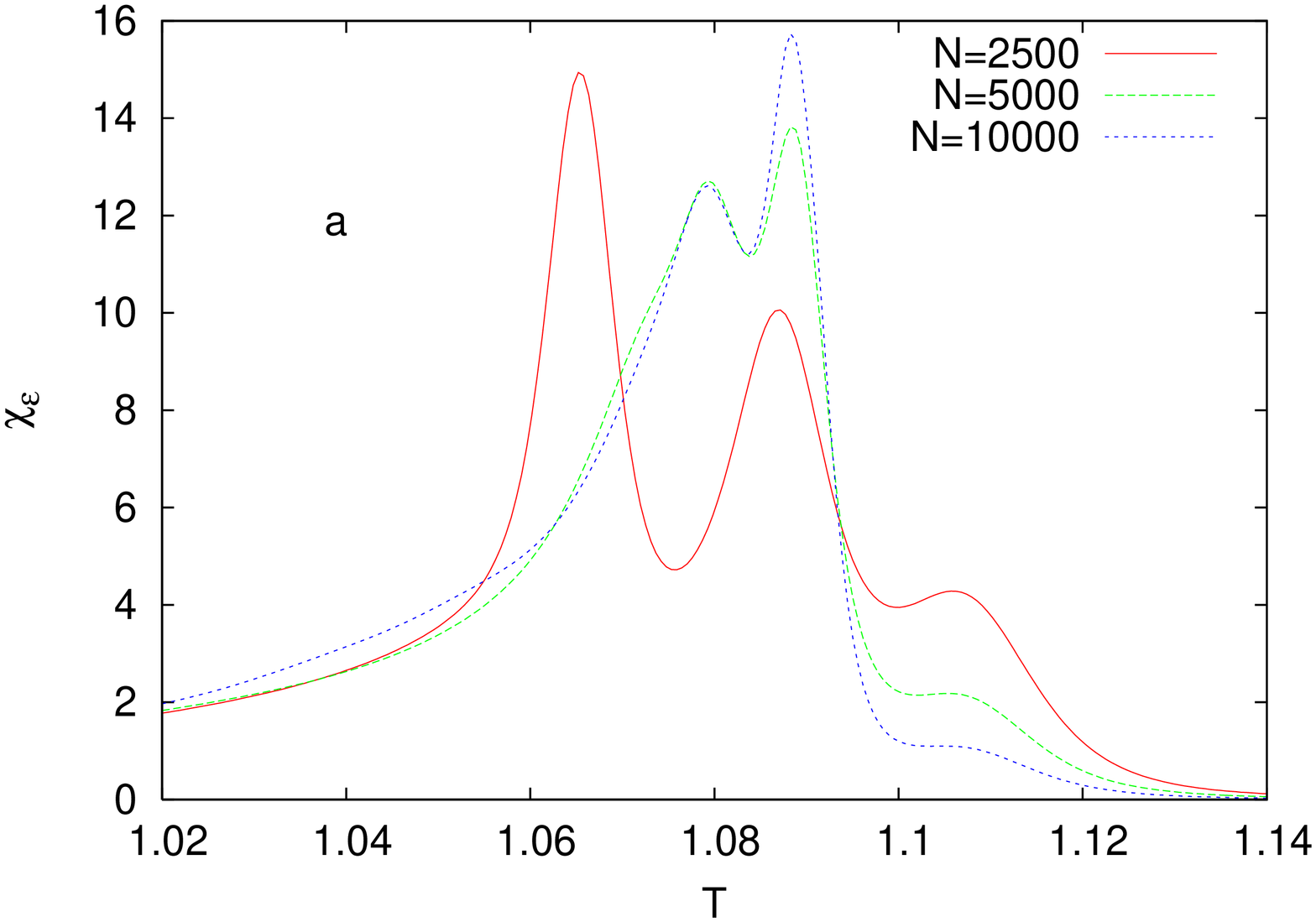,angle=270,width=8cm}
\epsfig{figure=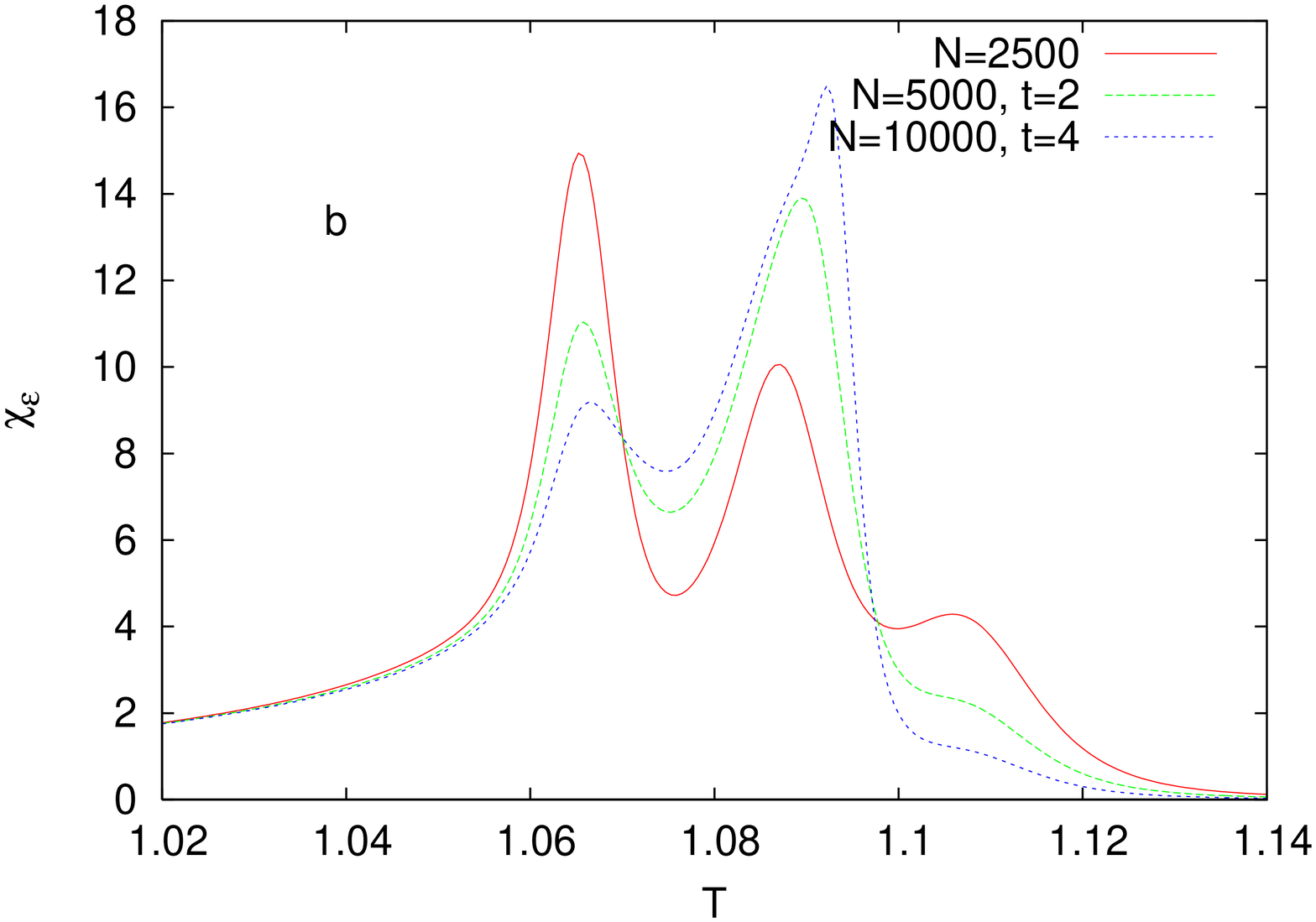,angle=270,width=8cm}
\caption{Susceptibility as function of temperature, plotted for given sequences
of lengths $N=2500$, $N=5000$ and $N=10000$, respectively. The randomly 
generated sequence of length $N=2500$ is the same in the two plots, 
corresponding to the one used in the previous figures, hence
the corresponding susceptibility is in particular the same than 
in [Fig. \ref{rev_new_fig2}b]. a) The sequence of length $N=5000$ coincides 
with the original 
sequence of length $N=2500$ for the first half, with the second half being 
randomly generated and, similarly, the sequence of length $N=10000$ coincides 
with the sequence of length $N=5000$ for the first half, with the second
half being randomly generated; b) The sequences of length $N=5000$ 
and $N=10000$ correspond to the concatenation of $t=2$, respectively $t=4$, 
copies of the original sequence of length $N=2500$.}
\label{rev_new_fig4}
\end{center}
\end{figure}

Moreover, checking the behaviour of the order 
parameters for sequences 
obtained by concatenation of $t$ copies of the original sequence, we can 
observe in [Fig. \ref{rev_new_fig1}] 
that the multi-step behaviour becomes less 
evident for increasing chain lengths. To better 
understand the meaning of this observation, we consider, in 
[Fig. \ref{rev_new_fig3}] and [Fig. \ref{rev_new_fig4}], 
the evaluations of the susceptibility for different possible instances of 
sequences of increasing lengths, involving the original sequence of length
$N=2500$, for which this observable displays three distinct peaks. 
In detail: in [Fig. \ref{rev_new_fig3}] the 
sequences of lengths $N=5000$ and $N=10000$ are randomly generated, with 
accordingly no correlations between the three increasingly longer sequences;
in [Fig. \ref{rev_new_fig4}a], the sequences of length $2N$ correspond to the 
concatenation of the original sequence of length $N$ with a randomly generated 
sequence of length $N$; in [Fig. \ref{rev_new_fig4}b] the sequences of length 
$tN$ correspond to the concatenation of $t$ 
instances of the original sequence of length $N$.  

Most strikingly, in all these instances, increasing values for $N$ do not
appear to be associated with the concomitant appearance of a larger number of 
peaks. Nevertheless, we can observe in these figures interesting 
qualitative differences between the three cases. In [Fig. \ref{rev_new_fig3}], 
the behaviour of 
susceptibility for the longest sequences do not appear to be correlated with 
that of the shortest one, with notably such behaviour changing completely 
between $N=2500$ and $N=5000$ and the position of the absolute maximum for the 
sequence of length $N=5000$ corresponding to the position of a minimum for the 
sequence of length $N=2500$. On the contrary, in [Fig. \ref{rev_new_fig4}], 
the presence of 
correlations between sequences of various lengths appear to be associated, not 
surprisingly, with concomitant correlations in the plots of the corresponding 
susceptibilities.

More specifically, it can be noticed in [Fig. \ref{rev_new_fig4}] that in 
case a), which is the closest to the conditions of the 
Monte-Carlo like numerical studies on the disordered on-lattice SAW DNA 
model \cite{Co}, significant correlations between susceptibility plots are 
observed for the longest sequences ($N=5000$ and $N=10000$), whereas in case 
b), which corresponds to the case of sequences generated by concatenations of 
a same original sequence, strong correlations between susceptibility plots are
observed for all sequence lengths. In case b) it can be further 
noted that, with increasing chain lengths, the position 
of the absolute maximum for the susceptibility 
appears to approach the position for which crossing of the order parameters is 
observed in [Fig. \ref{rev_new_fig1}].  

As a matter of fact, the comparisons above highlight the 
importance of the protocols adopted for the generation of increasingly longer 
sequences. Moreover, in agreement 
with the picture proposed in \cite{CoYe}, and as further detailed below, these 
comparisons also suggest the importance of the 
contribution of rare regions in the sequences to the thermodynamic 
limit behaviour of the disordered system considered. Indeed, in the 
concatenation scheme, as can be noticed
notably in [Fig. \ref{rev_new_fig4}b], the maximum length 
for such rare regions is strictly restricted by the maximum length of rare 
regions in the original sequence used. 

\subsection{The scaling behaviours of $T_c(N)$ and $\delta T_c(N)$}
\label{discussion1_two}
\begin{figure}[hptb]
\begin{center}
\leavevmode
\epsfig{figure=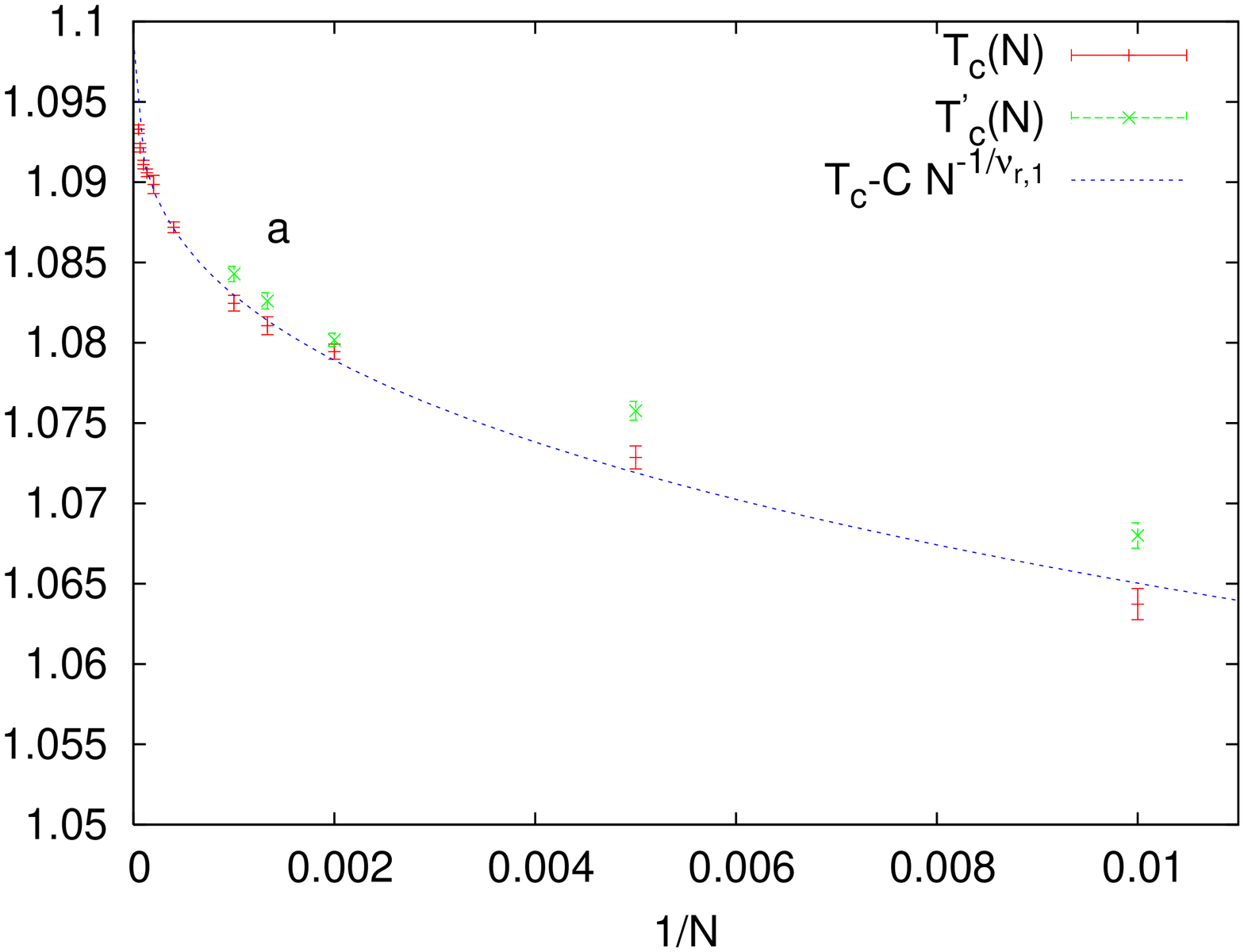,angle=270,width=8cm}
\epsfig{figure=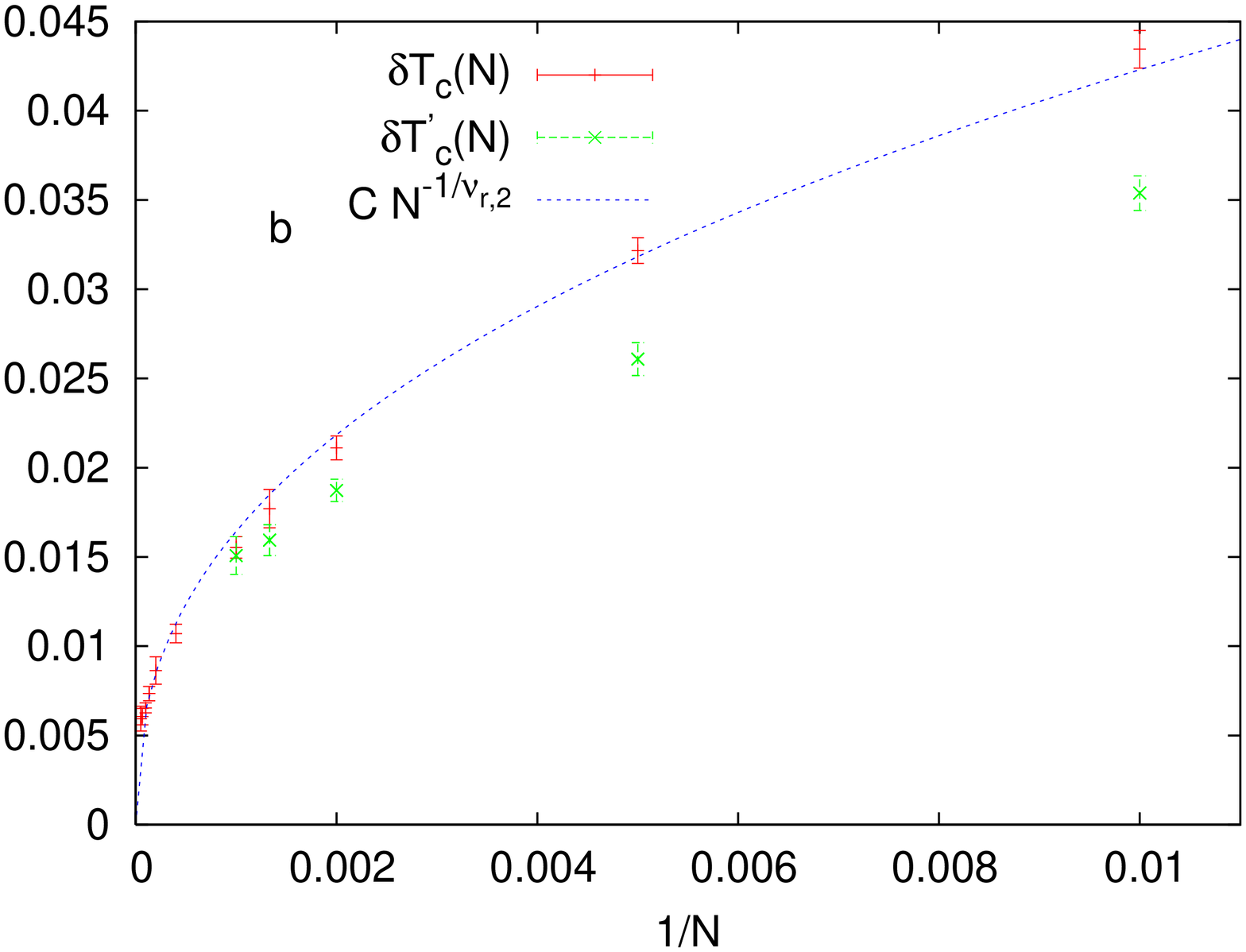,angle=270,width=8cm}
\caption{a) Plots of the mean values, $T_c(N)$ and $T_c'(N)$, of the 
pseudo-critical temperatures as functions of $1/N$. $T_c(N)$ values are plotted 
for the various chain lengths (red marks with error bars), corresponding
for a given $N$ to the mean value of $T_c(\{ \epsilon_i \},N)$, associated 
with the temperature for which the susceptibility reaches 
its absolute maximum, for the various sequences.
In addition, mean values of $T_c'(\{ \epsilon_i \},N)$ are plotted for 
$N \le 2500$ (green marks with error bars), defined as the crossing points 
of the order parameters $\theta(\{ \epsilon_i \},tN,T)$, for $t=1,2,4$.
The dotted line corresponds to the best fit of $T_c(N)$, 
according to the scaling law
$T_c(N) \simeq T_c+C N^{-1/\nu_{r,1}}$ (with $T_c=1.098 \pm 0.001$ and 
$1/\nu_{r,1}=0.33 \pm 0.03$). b) Plots of the fluctuations $\delta T_c(N)$ and 
$\delta T_c'(N)$, associated with $T_c(N)$ and $T_c'(N)$ in a), as functions
of $1/N$. The dotted line corresponds to the best fit for $\delta T_c(N)$, 
according to the scaling law  
$\delta T_c(N) \propto N^{-1/\nu_{r,2}}$ (with $1/\nu_{r,2}=0.405 \pm 0.01$).
}
\label{rev_new_fig5}
\end{center}
\end{figure}

The values of $T_c(N)=\overline{T_c(\{ \epsilon_i \},N)}$ and
$\delta T_c(N) = \{ \overline{[T_c(N)-T_c( \{ \epsilon_i \},N)]^2} \}^{1/2}$, 
as functions of $1/N$, are plotted in [Fig. \ref{rev_new_fig5}], along with the 
best fits obtained following $T_c(N) \propto  T_c + C \: N^{-1/\nu_{r,1}}$ and 
$\delta T_c(N) \propto N^{-1/\nu_{r,2}}$, hence allowing the possibility, in 
principle, of two different exponents 
(see Eqs.~(\ref{pseudotcirrev})-(\ref{pseudotcrev}) and the associated 
discussions).
In addition, data for 
the mean value and for the fluctuations of $T_c'(\{ \epsilon_i \},N)$, 
defined according to \cite{GaMo,MoGa}, are also plotted in 
[Fig. \ref{rev_new_fig5}] for $N \le 2500$.
For each $N$-value, we also
checked that the behaviour of $T_c(\{ \epsilon_i \},N)$ agrees with a Gaussian 
distribution.
We notice first of all that, interestingly, 
whereas strong finite size corrections are observed in the 
behaviour of average quantities when applying standard scaling laws 
\cite{CoYe}, it appears that data on the mean value 
and on the fluctuations of the 
pseudo-critical temperature agree well with 
the corresponding expected laws on the whole $N$-range considered.  

On such basis it is then straightforward 
to determine whether the values of the 
exponents are different, as associated to typical and average quantities, 
respectively.  Indeed, it is immediately 
obvious from the figure that the data for $T_c(N)$ and $\delta T_c(N)$ (as the 
ones for $T'_c(N)$ and $\delta T'_c(N)$) display essentially the same 
$N$-dependence.  This result implies that the pseudo first order
transition scenario cannot describe the observed behaviour in the present case, 
since the  mean value and the fluctuations of the pseudo-$T_c$
are ruled by the same exponent $\nu_r$ (which appears to be larger 
than $2$).

\begin{table}
\begin{center}
\begin{tabular}{||c||c|c||c||}
\hline
\hline
$N_0$ & $T_c$ & $1/\nu_{r,1}$ &$1/\nu_{r,2}$ \\
\hline
\hline
100 & 1.098$\pm$0.001 & 0.33 $\pm$ 0.03 & 0.405 $\pm$ 0.01 \\
\hline
200 & 1.101$\pm$0.002 & 0.27 $\pm$ 0.04 & 0.38 $\pm$ 0.01 \\
\hline
500 & 1.101$\pm$0.003 & 0.27 $\pm$ 0.06 & 0.37 $\pm$ 0.01 \\
\hline
750 & 1.099$\pm$0.003 & 0.32 $\pm$ 0.08 & 0.365 $\pm$ 0.01 \\
\hline
1000 & --- & --- & 0.34 $\pm$ 0.02 \\
\hline
2500 & --- & --- & 0.33 $\pm$ 0.06 \\
\hline
\hline
\end{tabular}
\caption{Estimations of $T_c$, $1/\nu_{r,1}$ and $1/\nu_{r,2}$ obtained 
disregarding chains of length $N < N_0$, for the different $N_0$ values 
considered. For $T_c$ and $1/\nu_{r,1}$ estimations are from $T_c(N)$ data, and 
for $1/\nu_{r,2}$ from $\delta T_c(N)$ data (see text).}
\end{center}
\label{nur1r2}
\end{table}

For further deepened analysis, [Tab. 5] reports estimations of $T_c$ and 
$1/\nu_{r,1}$ (from $T_c(N)$), and of $1/\nu_{r,2}$ (from $\delta T_c(N)$), 
obtained disregarding chain lengths $N < N_0$, for different $N_0$ values. 
This analysis does not reveal any obvious finite size corrections to scaling 
in the behaviour of $T_c(N)$. On the other hand $1/\nu_{r,2}$ appears to 
display a weak dependence on $N_0$, decreasing to the value 
$1/\nu_{r,2} \sim 0.33 \div 0.34$ when only chain lengths $N \ge 1000$ are 
considered. Thus we are led to similar values for the two exponents, 
with $1/\nu_{r,1}=1/\nu_{r,2}=1/\nu_r=0.35 \pm 0.05$,  
and accordingly to the estimates: $\nu_r=2.9 \pm 0.04$, and 
$c_r=1+1/\nu_r=1.35 \pm 0.05$. 
It is moreover possible to notice that analysis of the mean value 
and of the fluctuations of 
$T_c'(\{ \epsilon_i \},N)$ leads to estimates 
($1/\nu_{r,1}=0.44 \pm 0.06$ and $1/\nu_{r,2}=0.38 \pm 0.01$) which are 
consistent with the previous ones, even though in this case only chain lengths 
$N \le 2500$ are considered. 

Pseudo-critical temperatures appear thus to represent particularly interesting 
observables for the model here, 
with their scaling behaviour clearly in accordance with the 
typical scenario 
corresponding to relevance of disorder,   
the same than the one in the analysis of random 
ferromagnets \cite{AhHa,WiDo}: the numerical analysis provides 
evidence for a smooth phase transition (as already found in 
\cite{CoYe}), 
the thermodynamic limit behaviour being ruled by a single
correlation length, in agreement notably with the 
mathematical 
findings in \cite{GiTo1,GiTo2,Gi,To}. 
More precisely, our present best estimation of the random 
critical point correlation length exponent ($\nu_r=2.9 \pm 0.04$)  
is perfectly consistent with the previous 
evaluation in \cite{CoYe}, obtained through the 
scaling of the maximum of the specific heat averaged over disorder in the 
standard way. 

Importantly, it is also clear that, for the pseudo-critical temperature 
oriented analyses, asymptotic behaviours appear to be reached for 
chain lengths shorter than those  
considered in the present study (up to $N = 2 \cdot 10^4$), despite the strong 
finite size 
corrections to scaling characterizing the behaviour of the various 
thermodynamic observables in this disordered 
model \cite{Co,CoYe}, 
This situation is expected to be 
related to the choice of parameters \cite{CoYe},
and in order to reach a quantitative description of 
the dependency of the behaviour of the model on these parameters it 
would be relevant to determine 
a crossover chain length $N^*$, below which 
the present data could also agree with a pseudo first 
order transition. 

With this respect, in order to grasp the behaviour of the model
in the thermodynamic limit, looking in detail to [Fig. \ref{rev_new_fig5}], 
it appears 
important on qualitative bases to have $N > 1000$ ({\em i.e.} $1/N < 0.001$). 
Indeed, with $N \le 1000$, the average $T_c(\{ \epsilon_i \},N)$, as a 
function of $1/N$, would appear instead to be  
adequately fitted with a straight line, while this is not true for 
$\delta T_c(\{ \epsilon_i \},N)$. Thus the behaviour of these observables 
would be in agreement with a transition with $\nu_{r,2}> \nu_{r,1} = \nu_p=1$. 
Accordingly, $N^* \sim 1000$ can be retained as the evaluation of 
a crossover chain length suggested by data on the mean value 
and on the fluctuations of the 
pseudo-$T_c$ in our case. Clearly, such value corresponds to an evaluation 
{\em from below}, as the considered observables appear to be the less affected 
by corrections to scaling. 

\subsection{Comparisons between different ways of averaging}
\label{discussion1_three}
\begin{figure}[hptb]
\begin{center}
\leavevmode
\epsfig{figure=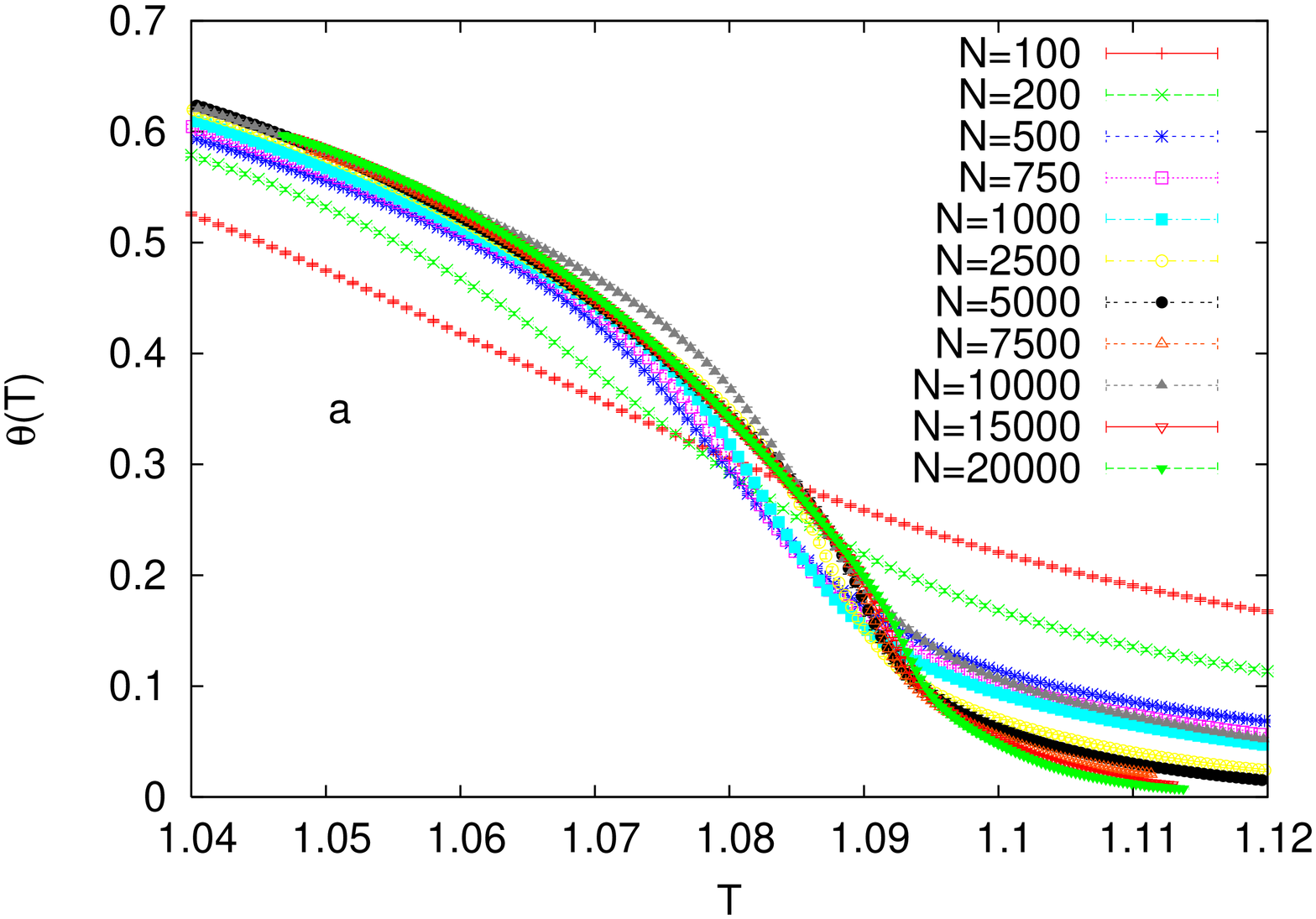,angle=270,width=8cm}
\epsfig{figure=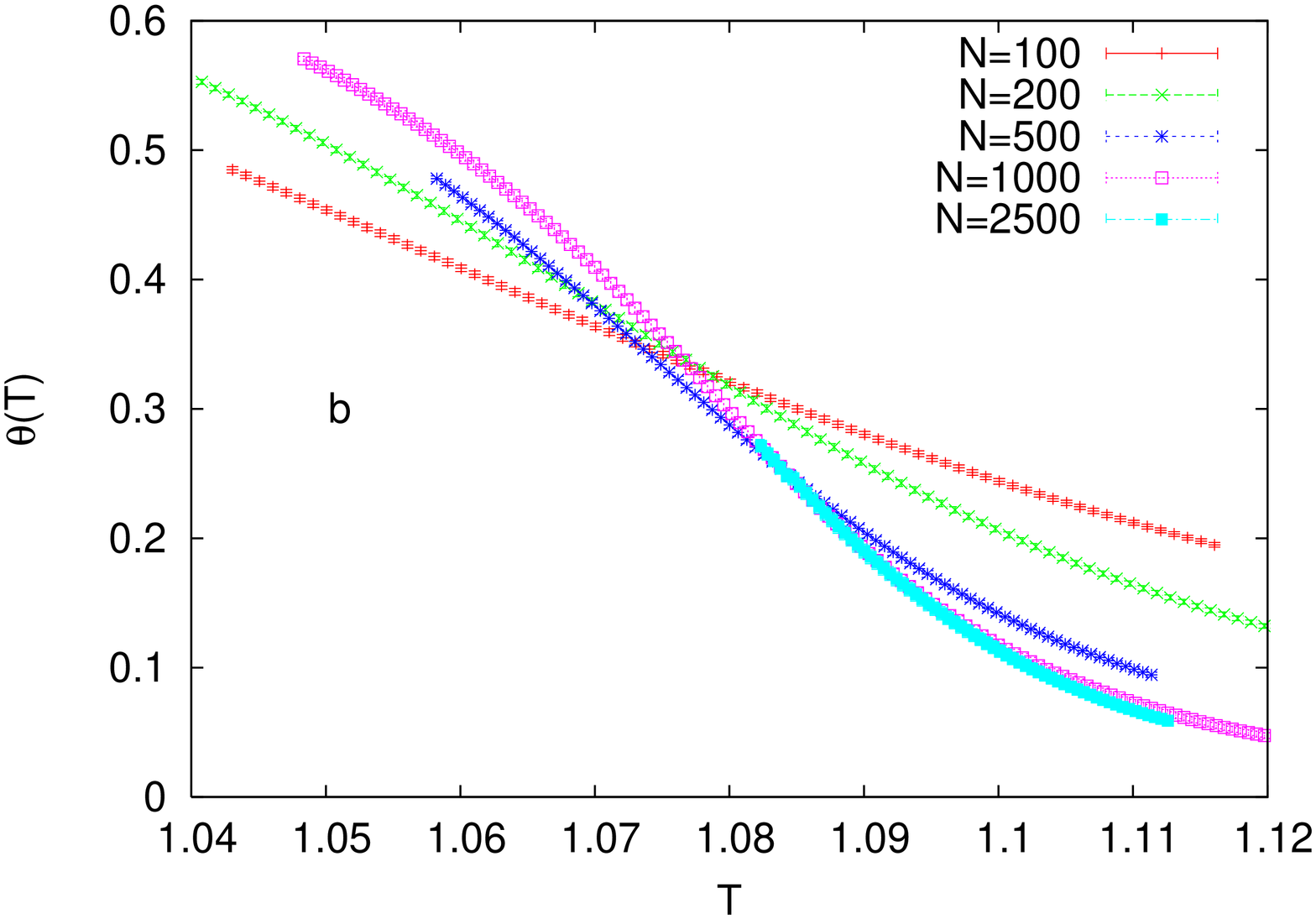,angle=270,width=8cm}
\caption{Plots of $\overline{\overline{\theta(\{ \varepsilon_i \},N,T)}}$: 
a) the pseudo-critical temperature is defined as the
value for which the absolute maximum of the susceptibility
is reached; b) for chain lengths $N \le 2500$, the pseudo-critical temperature 
is defined as the crossing point of the plots of
$\theta(\{ \varepsilon_i \},tN,T)$ (for $t=1,2$ and $4$, obtained from the 
concatenation of $t$ copies of the original sequence).}
\label{rev_new_fig6}
\end{center}
\end{figure}

The behaviour of $\overline{\overline{\theta( \{ \varepsilon_i \},T,N)}}$ is 
plotted in [Fig. \ref{rev_new_fig6}], following the two definitions of 
pseudo-critical temperatures that we considered. 
Indeed, the present analysis aims
to understand at a qualitative level the general results on quantities 
averaged by taking 
into account the pseudo-$T_c$ . With this
respect the plots in [Fig. \ref{rev_new_fig6}] 
can be compared to those presented in \cite{CoYe}, in which the 
standard averaging method was used. In any case, we 
do not perform any finite size scaling analysis for the order parameter
as, independent of the averaging 
method, 
the behaviour of this observable is not in 
agreement with the expected scaling law (hence with 
$N^{1-1/\nu_r} \tilde{\theta}[(T_c-T)N^{1/\nu_r}]$, 
with $\tilde{\theta}$ 
a scaling function, which can be derived from 
Eq.~(\ref{scaling}) with $e_r=\beta_r=\nu_r-1$). This feature was also observed 
both in \cite{GaMo} and in our previous work \cite{CoYe}.

From a qualitative point of view,
the data displayed in [Fig. \ref{rev_new_fig6}] show 
unambiguously that, with varying $N$, the order parameters do not 
cross at the same temperature. This situation stands in sharp contrast with 
the one characterizing
the pure model and also with the results reported in \cite{GaMo,MoGa}. 
As a matter of fact, this feature, clearly in
disagreement with the possibility of a pseudo first order character
for the behaviour of the model, was already observed in \cite{CoYe}. With the
analyses here it further appears that this result do not dependent on the 
averaging method used. In fact, this result concerning crossovers 
becomes even more obvious for shorter chain lengths upon 
looking at the behaviour of $\overline{\overline{\theta}}$. 
Finally, this observation 
holds independently of the definition of the pseudo-critical temperature, 
even though the positions of the crossing points appear to vary in fact still 
more rapidly with $N$ when averaging is performed by taking 
the pseudo-$T_c$ 
as the abscissa of the absolute maximum of the susceptibility. 
Thus, 
the analysis in terms of 
pseudo-critical temperatures appears to  
reduce, in general, the importance of the finite size effects, 
hence  
allowing notably to extrapolate
the correct thermodynamic limit behaviour from shorter chain lengths,
({\em i.e.} 
decreasing the effective $N^*$ value). 

For further clarification, we finally consider in detail the behaviour of the 
average susceptibility, whose evaluation is expected to be the most sensitive 
to different ways of averaging. Indeed, notably in the context of the study 
here, resorting extensively to the definition of the pseudo-critical temperature
as the position of the absolute maximum
of susceptibility for a given sequence, 
the quantity
${\overline{\overline{\chi(\{ \varepsilon_i \},T,N)}}}$ is
expected to best highlight, with its possible divergence, 
a pseudo first character in the behaviour of the model 
(see Eq.~(\ref{chimax})). 
Moreover, noticeably, the maximum of this quantity is expected to behave as a 
typical quantity, in the sense to be the observable the less affected by 
fluctuations in 
the pseudo-$T_c$ itself (see Eq.~(\ref{chimax})). From this point of 
view the study of this quantity is also expected to best highlight possible 
differences between typical and average 
behaviours.

\begin{figure}[htpb]
\begin{center}
\leavevmode
\epsfig{figure=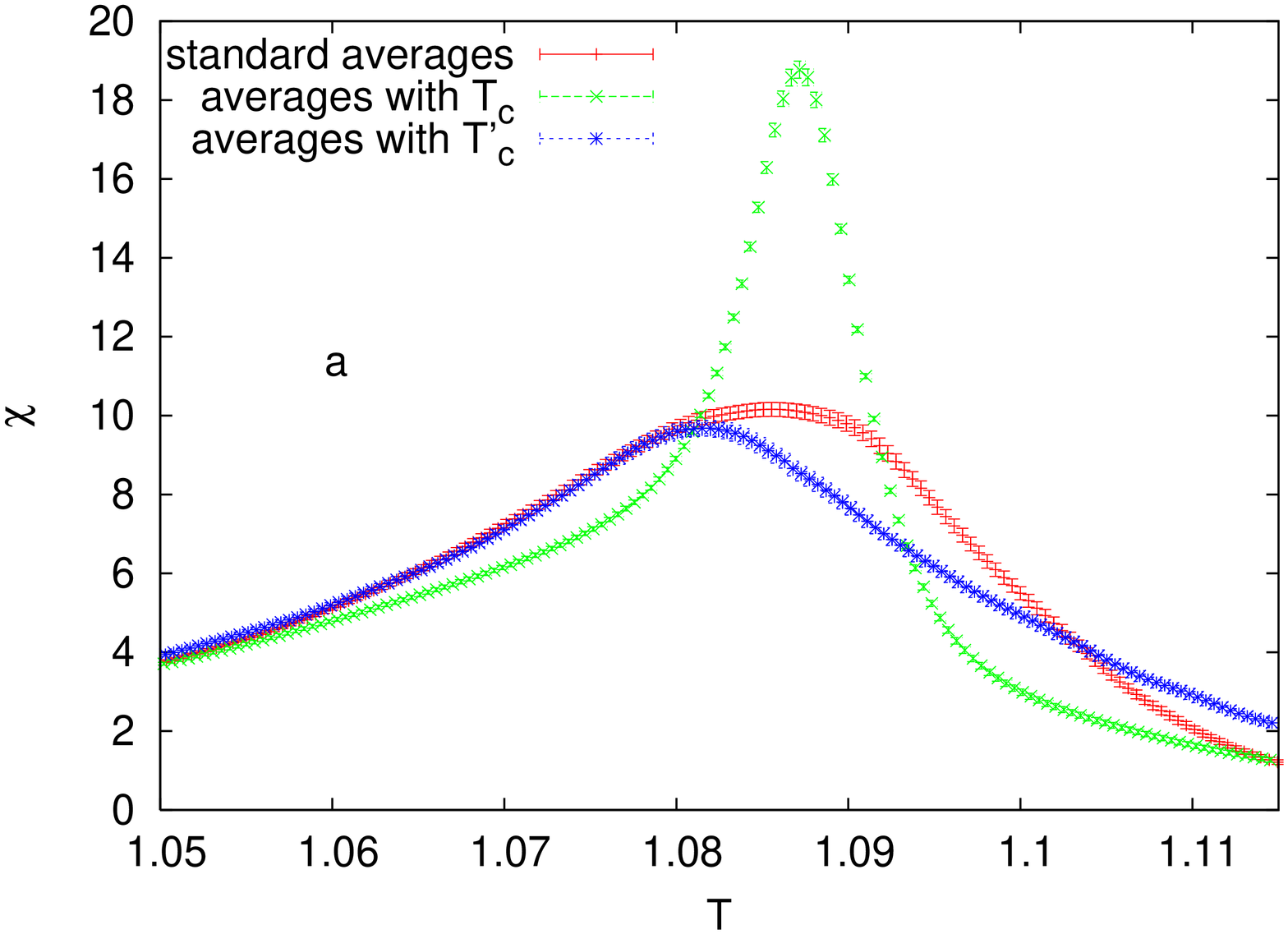,angle=270,width=8cm}
\epsfig{figure=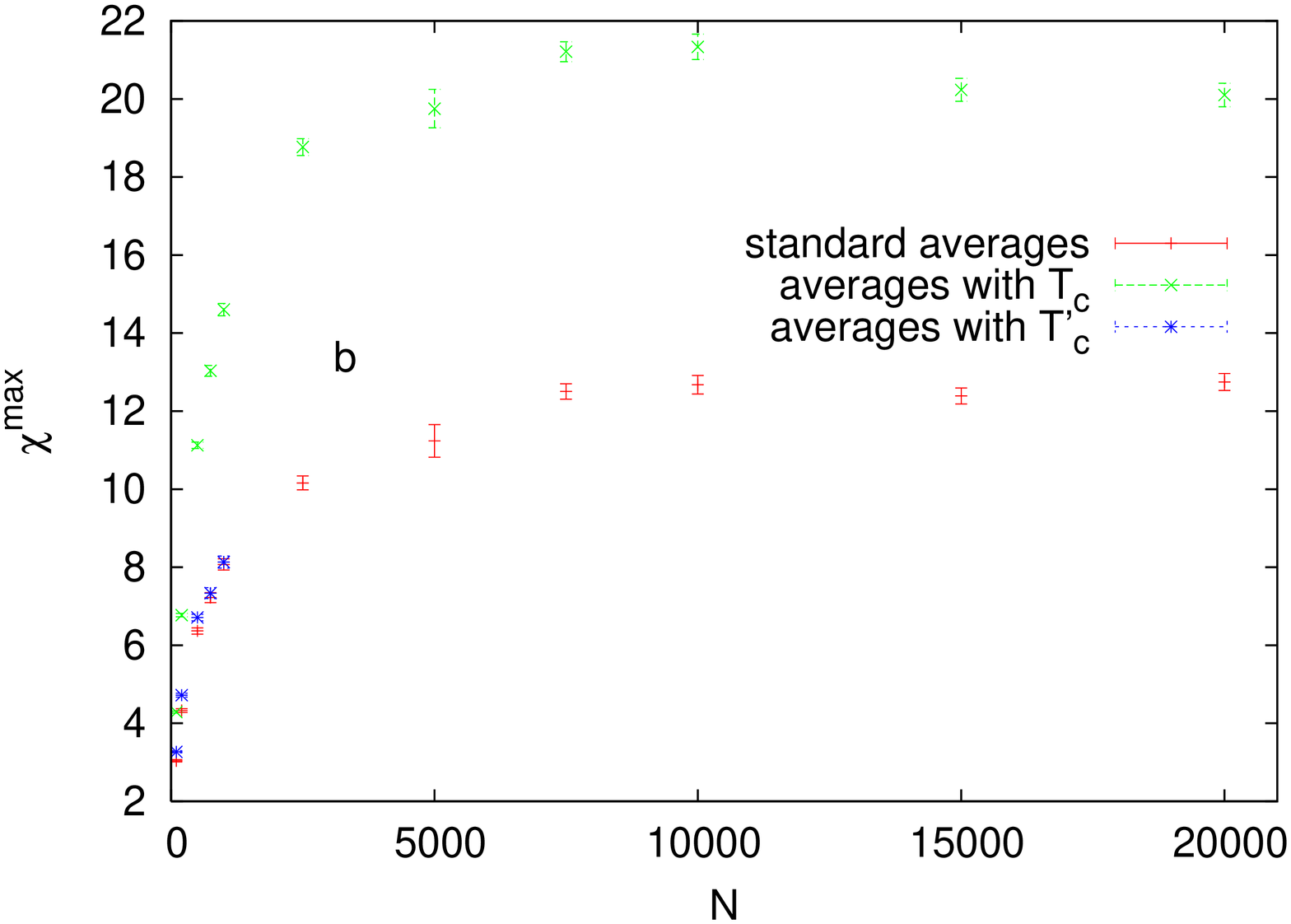,angle=270,width=8cm}
\caption{a) Plots of $\overline{\chi(\{ \varepsilon_i \},T,N)}$ and 
${\overline{\overline{\chi(\{ \varepsilon_i \},T,N)}}}$ for chain length 
$N=2500$ as function of the temperature. b) Plots of the corresponding 
maxima as function of the system size. In the first case averaging is 
performed
in the standard way (same data as in \cite{CoYe}). In the second case
averaging is performed taking into account the pseudo-critical temperature, 
following the two definitions considered ({\em i.e.} with 
$T_c(\{ \varepsilon_i \},N)$ the temperature where, for the given sequence,  
susceptibility reaches its absolute maximum and with 
$T_c'(\{ \varepsilon_i \},N)$ the temperature corresponding to the 
crossing point of the order parameter, for the given sequence, with those 
of the sequences obtained by concatenating variable number of copies of the 
given sequence).}
\label{rev_new_fig7}
\end{center}
\end{figure}

In detail, for the susceptibility data the results, 
for $N=2500$, are plotted in [Fig. \ref{rev_new_fig7}a] for 
$\overline{\chi(\{ \varepsilon_i \},T,N)}$ (as obtained from 
Eq.~(\ref{standardaverages}) and corresponding to the same data as in 
\cite{CoYe}) and for 
$\overline{\overline{\chi(\{ \varepsilon_i \},T,N)}}$
(as obtained from Eq.~(\ref{newaverages})), using the two definitions 
of the pseudo-critical temperature ($T_c(\{ \varepsilon_i \},N)$ and 
$T_c'(\{ \varepsilon_i \},N)$, respectively). 
It is then obvious from the figure that the possible pseudo first order 
character of the 
behaviour of the model is indeed suggested the most 
clearly by  data on $\overline{\overline{\chi}}$ evaluated by 
defining the pseudo-$T_c$ as the abscissa of the absolute maximum of 
susceptibility (for a given sequence), 
{\em i.e.} $T_c(\{ \varepsilon_i \},N)$,
with the average susceptibility reaching a definitely 
highest maximum in this case. The observation is in 
agreement with the expected result, 
thereby confirming that 
the considered $T_c(\{ \varepsilon_i \},N)$  
is particularly appropriate for assessing disorder relevance. Moreover, 
in this context it becomes easy to determine the behaviour of the model 
with respect to the typical scenario.

In fact, when looking at the scaling 
behaviours of the maxima values as functions of $N$ in 
[Fig. \ref{rev_new_fig7}b], a qualitative good agreement is found 
between the different ways considered for 
performing the averages.
In particular, in all cases the maximum of 
the average susceptibility displays a crossing to 
an $N$-independent regime for the largest considered chain lengths, clearly
showing that there is no difference in the behaviour of typical 
and average quantities. The data therefore further
support a smooth transition
(with $\gamma_r=\alpha_r\le 0$, and accordingly $\nu_r=2-\alpha_r \ge 2$). 
From a different point of view, 
these findings also clearly confirm qualitatively 
that the thermodynamic limit behaviour of the model
is described by a single correlation length, in agreement with the result
following which the mean value and the fluctuations of the 
pseudo-$T_c$ scale with the 
same exponent (namely, $\nu_r=2.9 \pm 0.4$).
It is worth recalling that in our previous work \cite{CoYe}, by
fitting the data for the maximum of the specific heat (averaged in the standard 
way) to the law $C_1-C_2 N^{\alpha_r/\nu_r}$, we obtained
(with the first two 
points disregarded) $\alpha_r/\nu_r= -0.3 \pm 0.1$, and hence 
$\nu_r=2.9 \pm 0.6$. Hence, it would be meaningless
to repeat the analysis 
on $\overline{\overline{\chi}}^{max}$ here. 

Inasmuch as the data the data for 
$\overline{\max_T \{\chi(\{ \varepsilon_i \},T,N)\}}$
display an abrupt change as function of $N$, it is all the more appropriate to 
introduce a crossover chain length $N^*$, for characterizing the
slow approach of the model to the asymptotic regime.
More precisely, we observe a shift from 
a short chain increasing behaviour to a long chain nearly
constant one, around a value of $N^*\sim 2500$, which is 
compatible both with the estimate from below ($N^* \sim 1000$)
obtained qualitatively from 
the scaling of the mean value and of the fluctuations of 
the pseudo-critical temperature 
and with the (more accurate) 
estimate from above ($N^{*} \sim 2.500 \div 5.000$) obtained 
according to the 
behaviour of the loop-length probability distribution (see below).

\subsection{Non self-averageness parameter related to susceptibility}
\label{discussion1_four}
As already recalled, both in the typical
case of disorder relevance and
within the pseudo first order transition scenario, one expects 
strong non self-averageness in the thermodynamic observables which are 
singular at the critical point. Accordingly, the parameter defined by 
Eq.~(\ref{nsa}), which measures the relative 
fluctuations of the observable averaged over disorder in the standard way, 
should display a constant behaviour instead of decreasing as a function of the 
system size $N$ at $T_c$.
In fact, 
numerical evidence was reported in \cite{MoGa} for such a 
behaviour of the non self-averageness parameter related to $\overline{\theta}$, 
for the disordered PS models considered with different $c_p$ values, and 
notably for the one with $c_p=2.15$. It is therefore reasonable to assume that 
the same result should hold in the present system.  
Indeed, intuitively enough, since the 
single sequence order parameter displays multi-step behaviour in a significant 
fraction of the samples, the sequence-to-sequence fluctuations of this 
observable, averaged in the standard way, should play a role at least of the 
same importance than in the model studied with $c_p=2.15$ in \cite{MoGa}, 
in which this behaviour is not observed. 

\begin{figure}[hptb]
\begin{center}
\leavevmode
\epsfig{figure=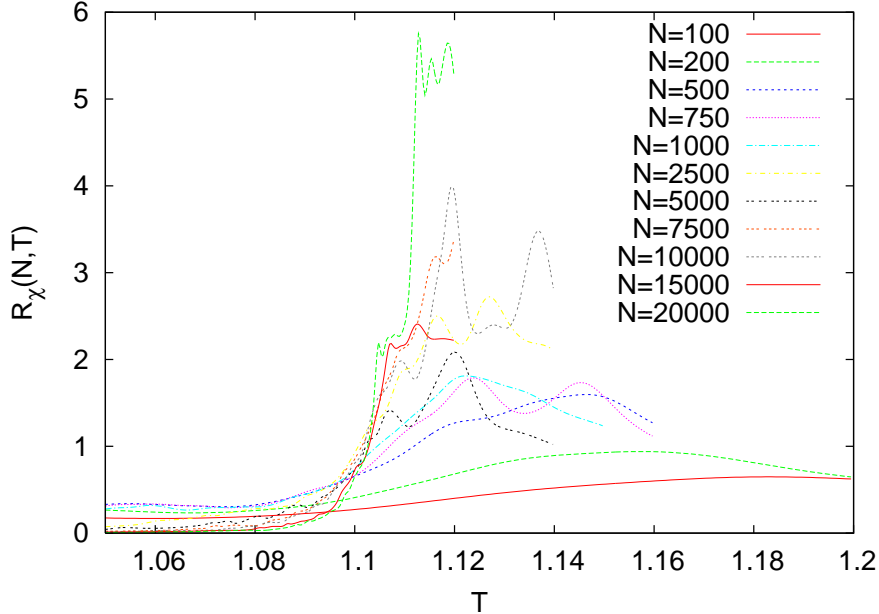,angle=270,width=12cm}
\caption{Plots of the non self-averaging parameter related to susceptibility, 
${\cal R}_{\chi}(N,T)$, as function of the temperature, for the
different chain lengths considered. No accurate evaluations of the errors 
are performed,
even though they are expected to be at most of the order of the 
observed oscillations of the quantity.} 
\label{rev_new_fig8}
\end{center}
\end{figure}

In this background, we  instead focus here on the
non self-averageness parameter related to 
the susceptibility, ${\cal R}_{\chi}(N,T)$, whose behaviour as function of
$T$ is plotted in [Fig. \ref{rev_new_fig8}], for the different sizes 
considered. 
First, letting aside the two shortest lengths $N=100$ and $N=200$, in the plots 
of ${\cal R}_{\chi}(N,T)$ no evident dependence on the 
chain length $N$ is observed for the heights of 
peaks, thus confirming the expected strong non self-averageness
of this observable in the present model. 
The plots in [Fig. \ref{rev_new_fig8}] display 
rather irregular behaviours in the whole high-$T$ region, 
even though it is in general expected that 
${\cal R} \sim 1/N$ both above and below the critical point. This observation
can be explained by the fact that here the disorder
couples only to the low-temperature phase, with the order parameter being zero 
in the thermodynamic limit for $T>T_c$, where the two DNA strands are only 
linked 
at the origin. It is then particularly difficult to evaluate correctly in this 
region the fluctuations due to disorder of the average 
susceptibility. 

Nevertheless, it is interesting to notice in [Fig. \ref{rev_new_fig8}], 
that the 
plots for different $N$-values display similar behaviours near the 
thermodynamic 
limit critical temperature $T_c \simeq 1.1$. In particular, a steep increase is 
observed in the plots immediately above $T_c \simeq 1.1$, with the steepness 
increasing with the size, and for temperatures approaching $T_c$ from above
for increasing sizes. In more quantitative terms, for the different 
$N$-values considered, the position of the highest ${\cal R}_{\chi}(N,T)$ peak, 
corresponding to the best evaluation of the abscissa of its absolute maximum as 
allowed by our present statistics, 
appears to be interpretable as the mean value
of a quantity behaving as a pseudo 
critical temperature (or in any event as an 
$N$-depending evaluation of the thermodynamic limit critical temperature
$T_c$), which we can denote $T^{\cal R}_c(N)$. 
In fact, we checked that 
the scaling law $T^{\cal R}_c(N)=T_c+C N^{-1/\nu_r}$ 
is also valid for this observable. With respect to the
previous cases, corresponding to the two different 
definitions of $T_c(\{ \varepsilon_i \},N)$, it is clear 
that the constant has the opposite sign, with
anyway the fit leading to compatible 
estimations of $T_c$ and $\nu_r$. 

\section{Characterization of the finite size behaviour: results and insights}
\label{discussion2}
\subsection{Crossover to the asymptotic regime}
\label{discussion2_one}  
The quantity:
\begin{equation}
P'( \{ \varepsilon_i \},N,T,l) \equiv (2l)^{c_p}P( \{ \varepsilon_i \},N,T,l),
\end{equation}
was introduced in our previous work \cite{CoYe}, as more appropriate
for capturing relevance of disorder, than the loop-length 
probability distribution $P( \{ \varepsilon_i \},N,T,l)$.
Indeed, at the critical point, the logarithm of 
$P'( \{ \varepsilon_i \},N,T,l)$ is expected to be not constant and 
proportional to $(c_p-c_r) \log l+C$ as soon as $c_r<c_p$. 

In more detail, we can consider 
both $\log \overline{P'( \{ \varepsilon_i \},N,T,l)}$ and
$\overline{\log P'( \{ \varepsilon_i \},N,T,l)}$, which should allow
to capture the behaviours of the average and typical correlation lengths 
respectively, following the picture put forward in \cite{GaMo,MoGa} (with in 
fact the second case better described as a mixed average). 
The behaviour of these quantities for $T \simeq T_c$, as shown in \cite{CoYe},  
was instead found compatible with a smooth transition, but
it is noticeable that the evaluated critical exponent $c_r$ 
appears to depend on $N$. In fact, such dependency
is more evident when considering the 
average quantity $\overline{\log P'( \{ \varepsilon_i \},N,T_c,l)}$,
further displaying more important deviations from the expected behaviour 
($\propto \log l$) than $\log \overline{P'( \{ \varepsilon_i \},N,T,l)}$. 

In summary, the analysis of the data in 
\cite{CoYe} suggested the asymptotic condition 
$c_r < 1.5$, which is confirmed by our present 
evaluation $c_r=1.35 \pm 0.05$ obtained from the pseudo-$T_c$ behaviour. 
Nevertheless, to further clarify the situation,  
attempting to better characterize the finite size
behaviour of the present model, we are led to study in detail  
both $\overline{\log P'( \{ \varepsilon_i \},N,T_c,l)}$ and
$\log \overline{P'( \{ \varepsilon_i \},N,T,l)}$ on the whole 
relevant $T$-range, by introducing effective ($N$-dependent) critical exponents 
$c_{r,1}(N)$ and $c_{r,2}(N)$, as well as correlation lengths 
$\xi_1(N,T)$ and $\xi_2(N,T)$. 

We found that the 
data concerning $\log \overline{P'( \{ \varepsilon_i \},N,T,l)}$ are
accurately described by the expected scaling law, which can be easily obtained
from Eq.~(\ref{plscaling}). Accordingly, the behaviour of 
$\log \overline{P'( \{ \varepsilon_i \},N,T,l)}$, which should be
ruled by the fluctuations of the pseudo-$T_c$, is in agreement with:
\begin{equation}
\log \overline{P'( \{ \varepsilon_i \},N,T,l)} 
\simeq (c_p-c_{r,2}(N)) \log l-l/\xi_2(N,T)+C(N).
\label{cr2xi2}
\end{equation}
On the other hand, in order to obtain a satisfactory fit (within 
the errors) for the data concerning 
$\overline{\log P'( \{ \varepsilon_i \},N,T,l)}$ (with such analysis 
expected to better capture the typical loop-length probability
distribution behaviour), it appears necessary to introduce, in addition, a 
quadratic contribution in $l$, {\em i.e.}:
\begin{equation}
\overline{\log P'( \{ \varepsilon_i \},N,T,l)} 
\simeq (c_p-c_{r,1}(N)) \log l-l/\xi_1(N,T)-C_1(N) l^2+C_2(N),
\label{cr1xi1}
\end{equation}
with the constant $C_1(N)$ tending towards zero roughly as $1/N$.

Such approach appears adequate to describe the strong finite size
corrections to scaling characterizing the present model, and 
above all it allows a clear definition of $N^*$.
In fact, the crossover between a short chain length regime in which the 
effective
exponents would be in agreement with a pseudo first order transition and a 
long
chain length one in which they would be compatible with the asymptotic values 
turns out to 
be quite abrupt. Therefore, one can obtain a quantitative evaluation of the
crossover chain length (which is an evaluation {\em from above}), as the 
length $N^*$ beyond 
which $c_{r,1}(N) \sim c_{r,2}(N) \sim c_r \simeq 1.35$ and 
$\lim_{T \rightarrow T_c^-}\xi_1(N,T) \sim \lim_{T \rightarrow T_c^-}\xi_2(N,T) 
\sim (T_c-T)^{-\nu_r}$, with $1/\nu_r=c_r-1 \simeq 0.35$. 

\begin{figure}[hptb]
\begin{center}
\leavevmode
\epsfig{figure=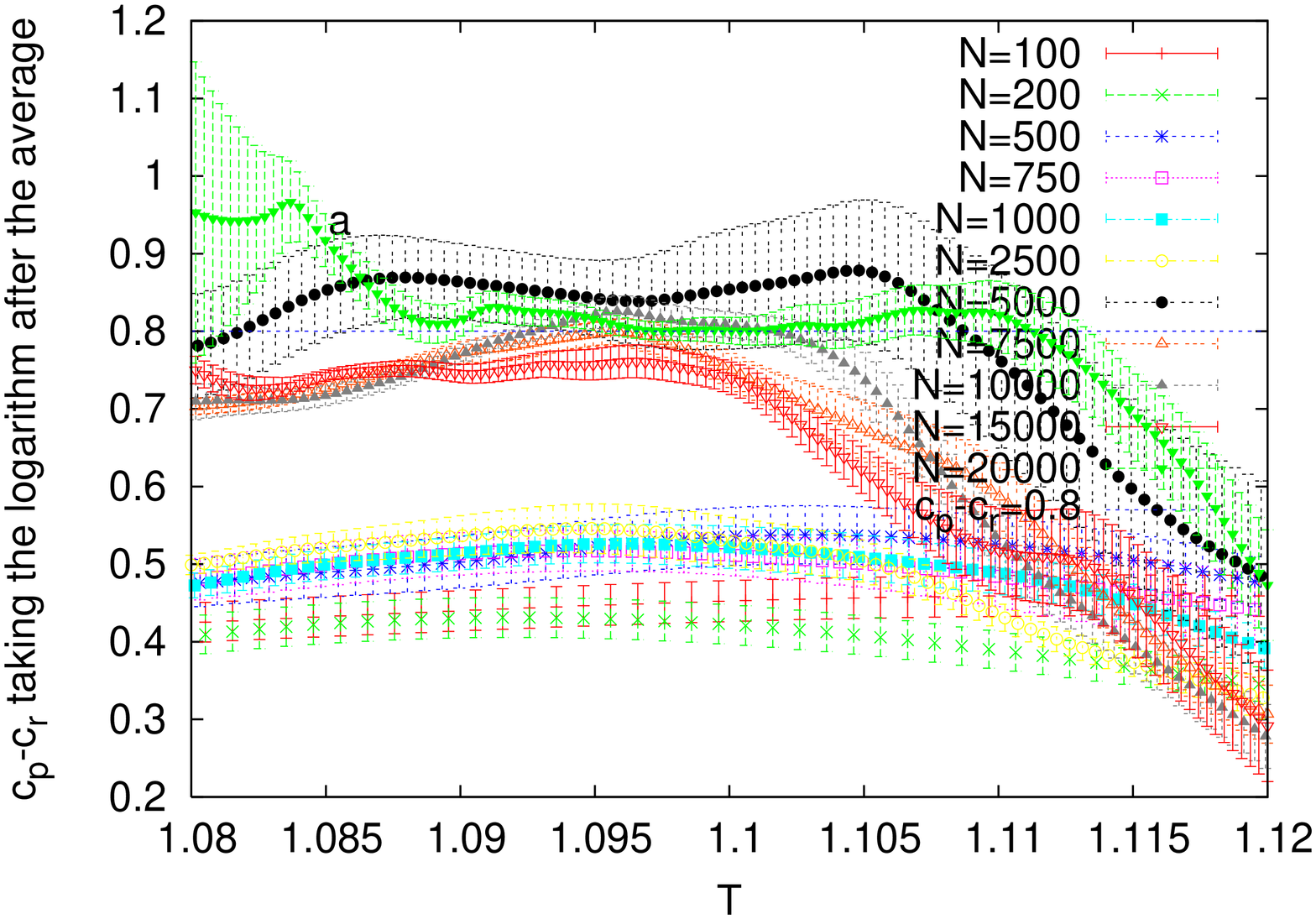,angle=270,width=8cm}
\epsfig{figure=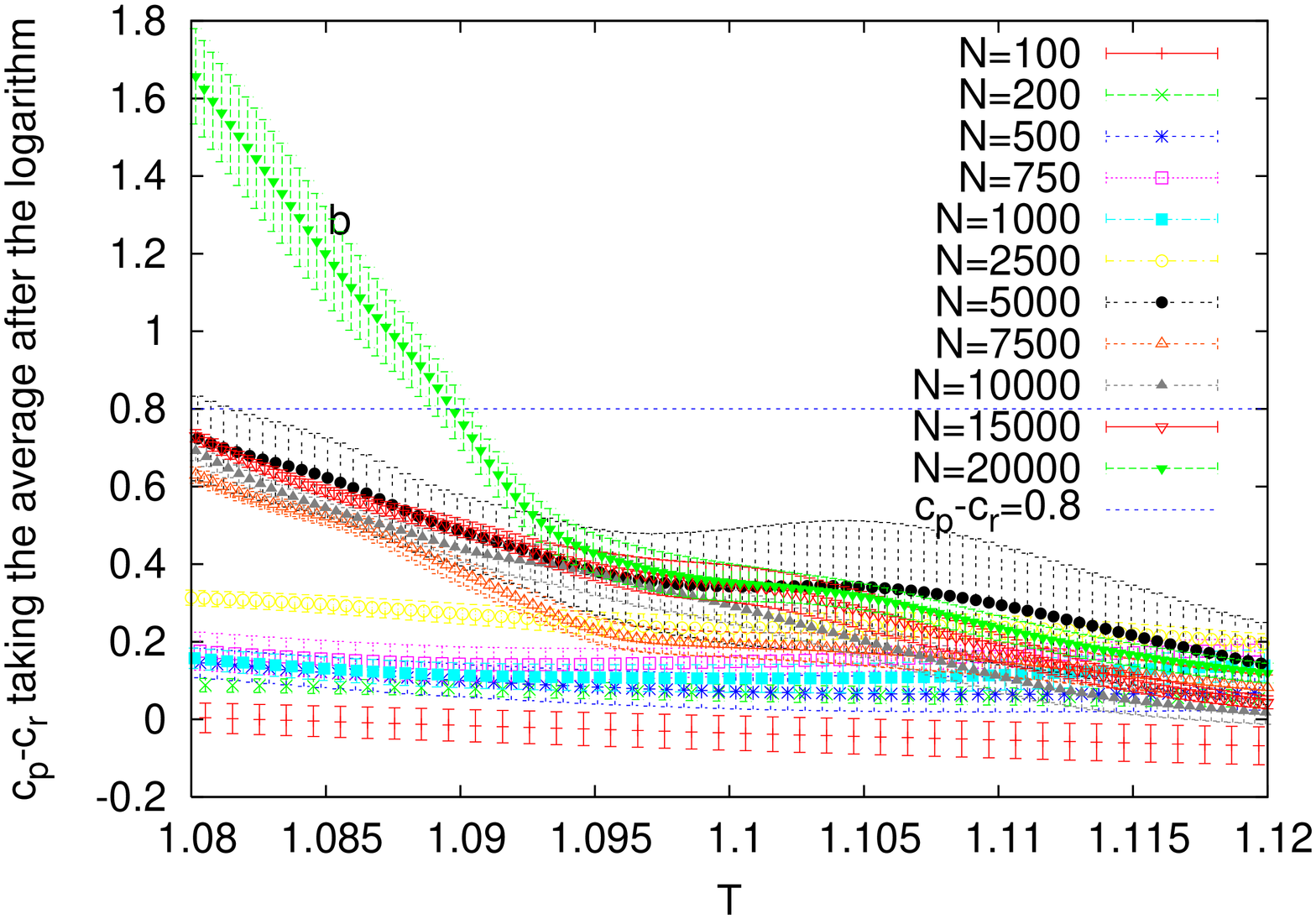,angle=270,width=8cm}
\caption{a) Evaluation, for the various chain lengths considered, 
of $c_p-c_{r,2}(N)$ from the fits (three parameters) of 
$\log \overline{P'( \{ \varepsilon_i \},N,T,l)}$ to Eq.~(\ref{cr2xi2}), 
as function of temperature. b) Evaluation, for the various chain 
lengths considered, of $c_p-c_{r,1}(N)$ from the fits (four parameters) 
of $\overline{ \log P'( \{ \varepsilon_i \},N,T,l)}$ to Eq.~(\ref{cr1xi1}), 
as function of temperature. 
For both analyses the errors are only of indicative value, as the results 
display some dependence on the $l$-range (with the range 
$l \in [3,N/3]$ considered corresponding to a reasonable choice), in 
particular for the shortest chain lengths. For comparisons, 
the expected asymptotic behaviour ($c_p-c_r \simeq 0.8$) is also represented in 
both panels.}
\label{rev_new_fig9}
\end{center}
\end{figure}

\begin{figure}[hptb]
\begin{center}
\leavevmode
\epsfig{figure=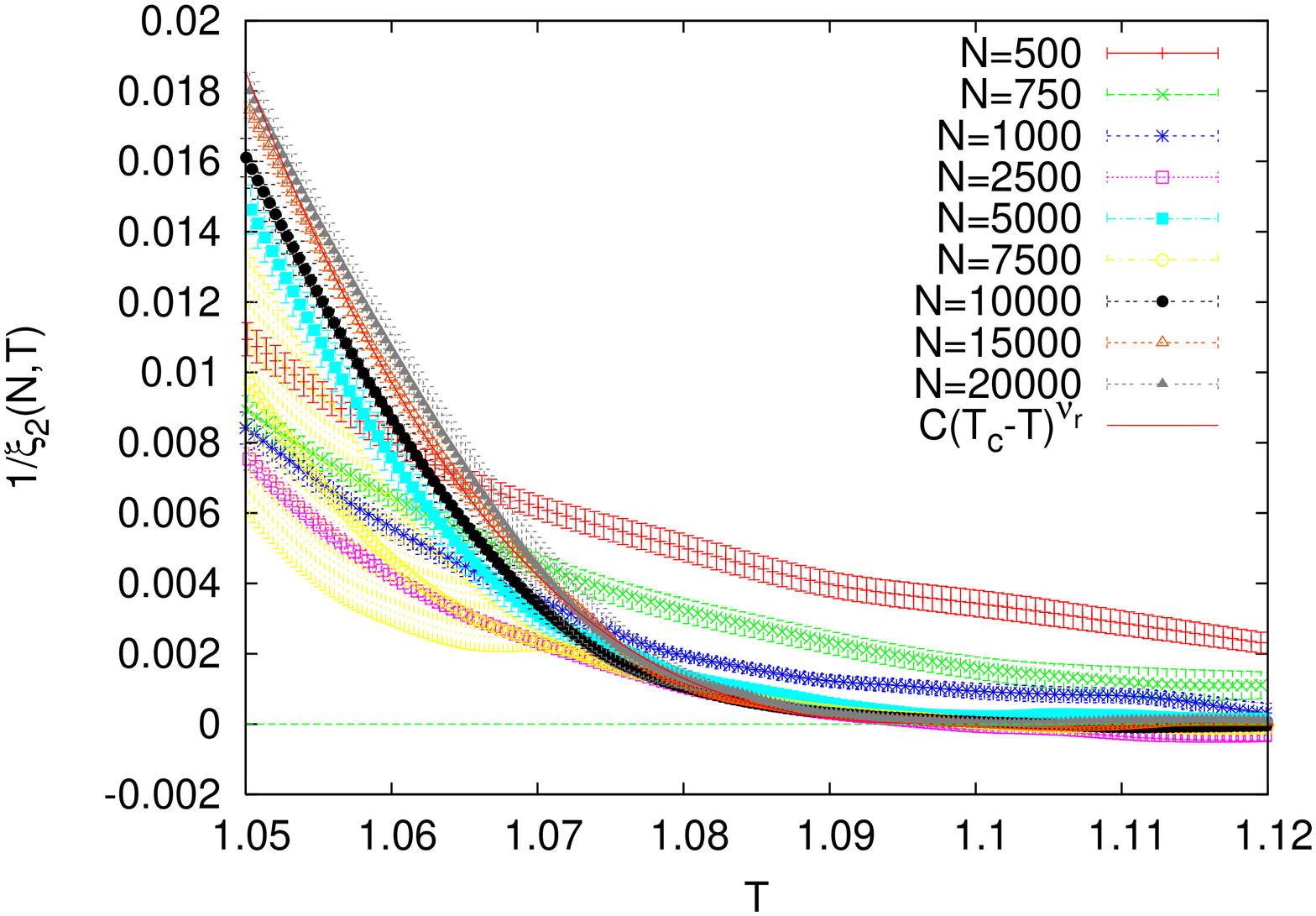,angle=270,width=8cm}
\epsfig{figure=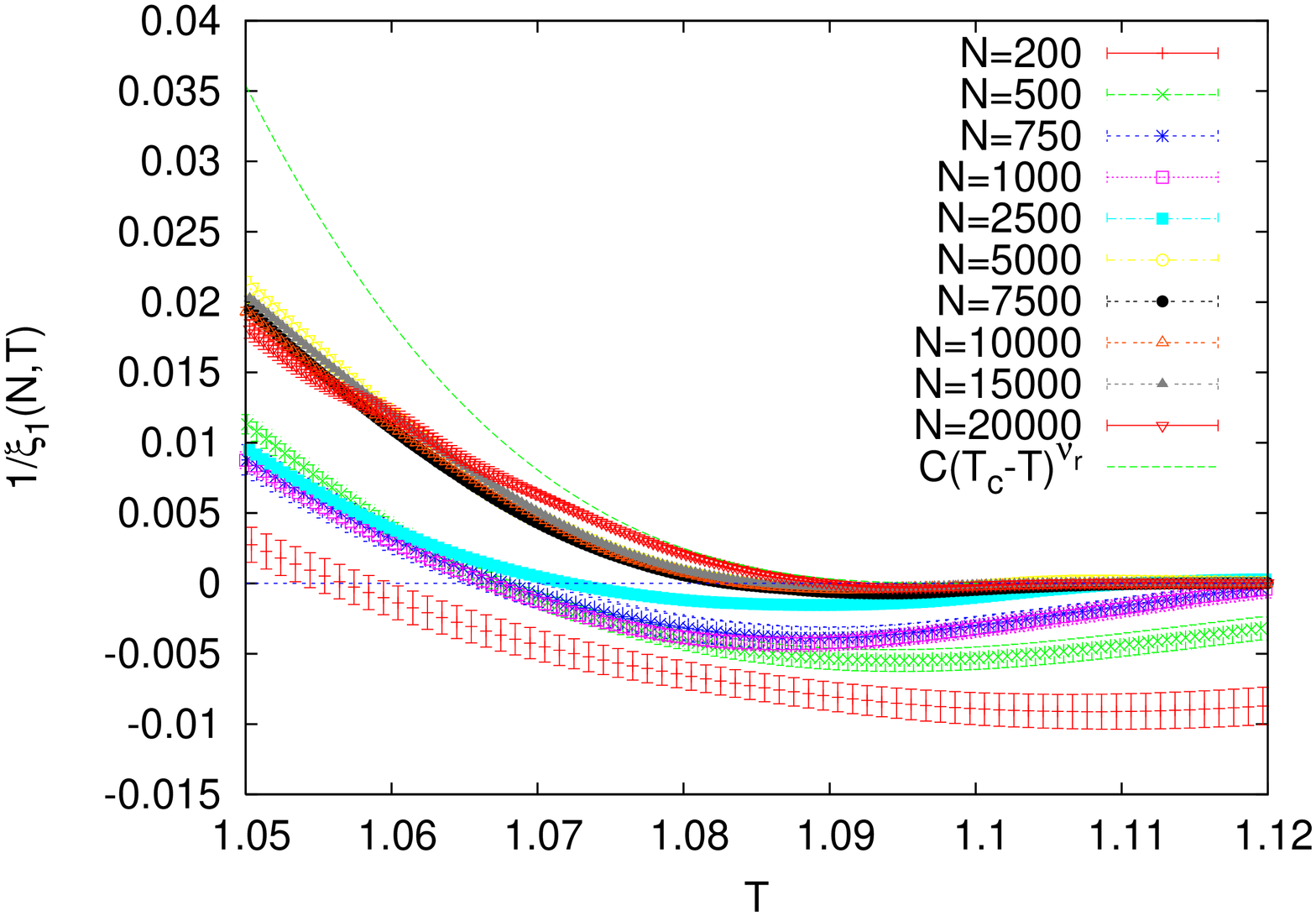,angle=270,width=8cm}
\caption{a) Evaluation of the inverse of the correlation length as 
function of temperature for the various chain lengths considered,
from the fits (three parameters) of 
$\log \overline{P'( \{ \varepsilon_i \},N,T,l)}$ to  
Eq.~(\ref{cr2xi2}). b) Evaluation of the inverse of the correlation length 
$\xi_1(N,T)$ as function of temperature for the various chain lengths 
considered, from the fits (four parameters) of
$\overline{ \log P'( \{ \varepsilon_i \},N,T,l)}$
to Eq.~(\ref{cr1xi1}). The same remarks as in [Fig. 9] apply to the 
significance 
of errors. For comparisons, the expected asymptotic behaviour 
($1/\xi(T) \propto (T_c-T)^{2.9}$) is also represented in both panels.}
\label{rev_new_fig10}
\end{center}
\end{figure}

The detailed results for the loop-length probability distribution exponents
and the inverse of the correlation lengths are presented in 
[Fig. \ref{rev_new_fig9}] and  [Fig. \ref{rev_new_fig10}], 
respectively, 
associated with Eq.~(\ref{cr2xi2}) and 
 Eq.~(\ref{cr1xi1}) in each of the two cases. 
More precisely, the 
figures plot the evaluations of $c_p-c_{r,2}(N)$, $c_p-c_{r,1}(N)$, 
$1/\xi_1(N,T)$ and $1/\xi_2(N,T)$, as obtained from 
the fits above, as functions of temperature for the different sizes considered. 
These plots also allow to compare the results with $c_p-c_r \simeq 0.8$ 
and $1/\xi(T) \sim (T_c-T)^{2.9}$ for $T \rightarrow T_c^-$ (the asymptotic 
inverse of the correlation length being zero in the whole high temperature 
phase). 

The plots in [Fig. \ref{rev_new_fig9}] and  [Fig. \ref{rev_new_fig10}] clearly 
illustrate the strong $N$-dependence of the quantities considered. Nonetheless,
as shown in particular in [Fig. \ref{rev_new_fig9}a], for chain lengths 
$N \ge 5000$, by allowing for a non-zero effective $1/\xi_2(N,T)$, 
$c_p-c_{r,2}(N)$ evaluations from 
$\log \overline{P'( \{ \varepsilon_i \},N,T,l)}$ are consistent with 
$c_r \simeq 1.35$ over a large temperature range around $T_c$. Accordingly, 
for the disordered PS model with $c_p=2.15$ studied, 
it appears that $N^*=2500 \div 5000$ can be assimilated to 
our best evaluation from above of the crossover length $N^*$. 
This conclusion also holds for the analysis of 
$\overline{\log P'( \{ \varepsilon_i \},N,T,l)}$ data, even though involving 
in this case more significant corrections. 

It is also very illustrative to notice 
that limiting the analysis to chain 
lengths $N < 2500$ can lead to interpretations significantly
different from those obtained considering the long chain length regime:
one would get from [Fig. \ref{rev_new_fig9}b] an estimate for 
$c_p-c_{r,1}(N)$ essentially compatible with the zero value
characteristic of the pure case ({\em i.e.} $c_{r,1}=c_p=2.15$ which 
would imply $\nu_{r,1}=\nu_p=1$), as well as (from 
[Fig. \ref{rev_new_fig9}a]) an estimate for $c_{r,2}(N)$ 
not significantly larger than 1.5 ({\em i.e.} $c_{r,2}=1+1/\nu_{r,2}$ 
with $\nu_{r,2}=2$). With this respect, the analysis here confirms, that in 
the regime below the crossover chain length $N^*$, it is difficult to 
extrapolate 
the correct thermodynamic limit behaviour on general grounds, with the results 
being indeed interpretable in such context
according to the picture implying a pseudo first order phase transition as 
proposed in \cite{GaMo,MoGa}.

\begin{figure}[hptb]
\begin{center}
\leavevmode
\epsfig{figure=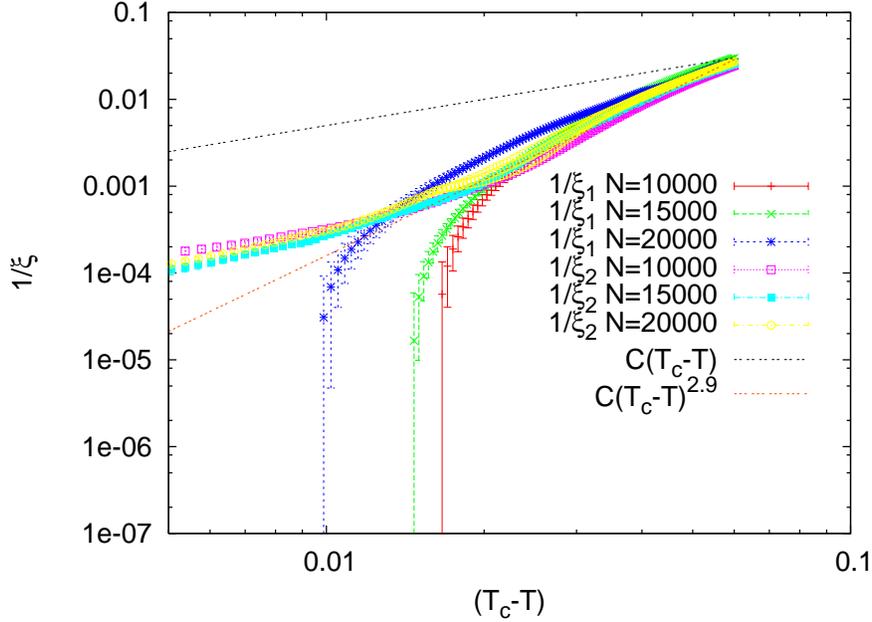,angle=270,width=12cm}
\caption{Plots of $1/\xi_2(N,T)$ and $1/\xi_1(N,T)$ on log-log scale as 
functions of $(T_c-T)$ (with $T_c \simeq 1.1$) for the three 
longest chain lengths considered ($N=10000$, $N=15000$ and $N=20000$; same data 
as in [Fig. \ref{rev_new_fig10}]). Data are plotted against the expected 
asymptotic behaviour of the inverse correlation length 
$1/\xi(T) \propto (T_c-T)^{\nu_r}$, with $T_c=1.1$ and $\nu_r=2.9$. 
For comparisons, the behaviour of $1/\xi(T) \propto (T_c-T)$ 
is also displayed.} 
\label{rev_new_fig11}
\end{center}
\end{figure}

\begin{figure}[hptb]
\begin{center}
\leavevmode
\epsfig{figure=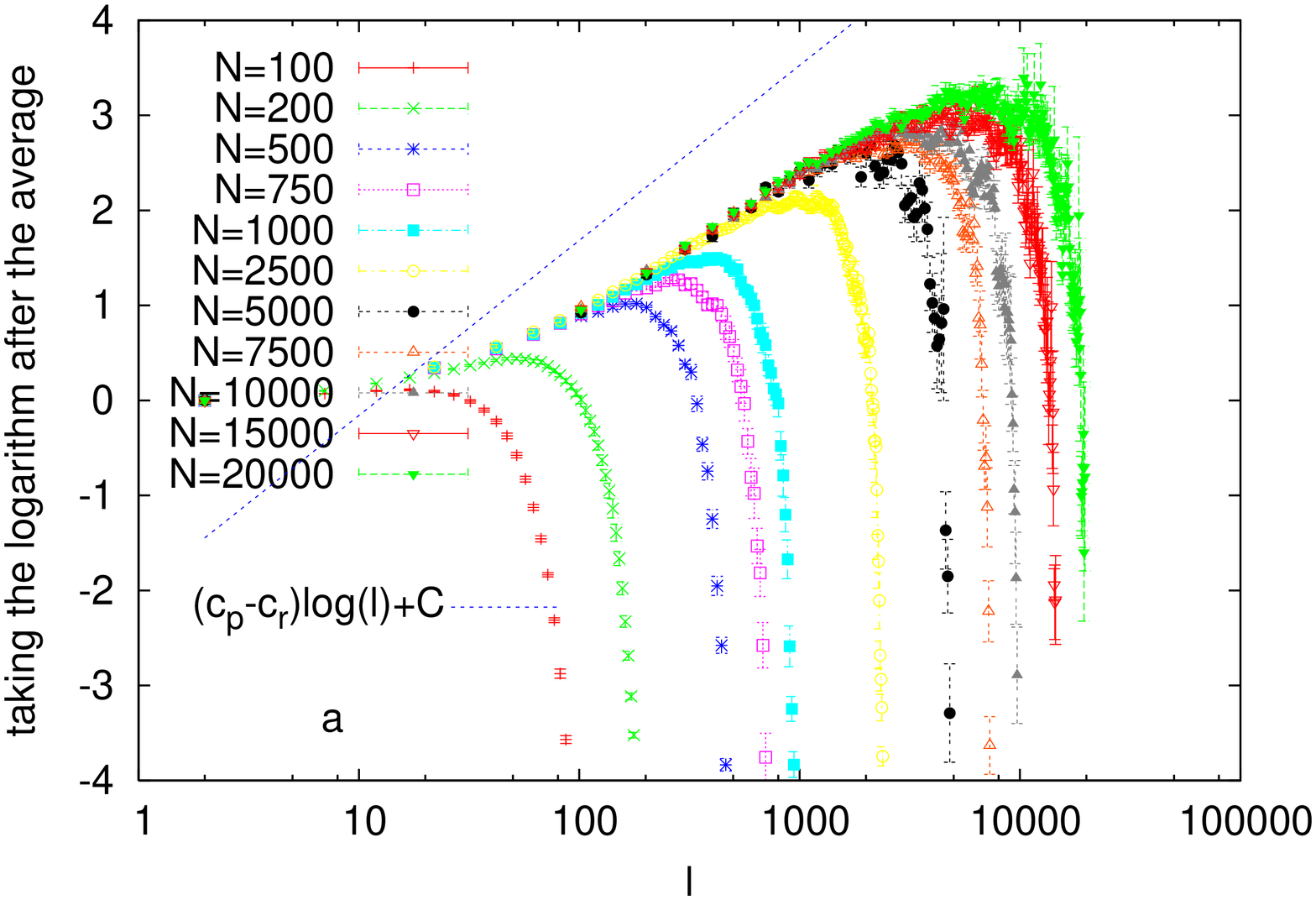,angle=270,width=8cm}
\epsfig{figure=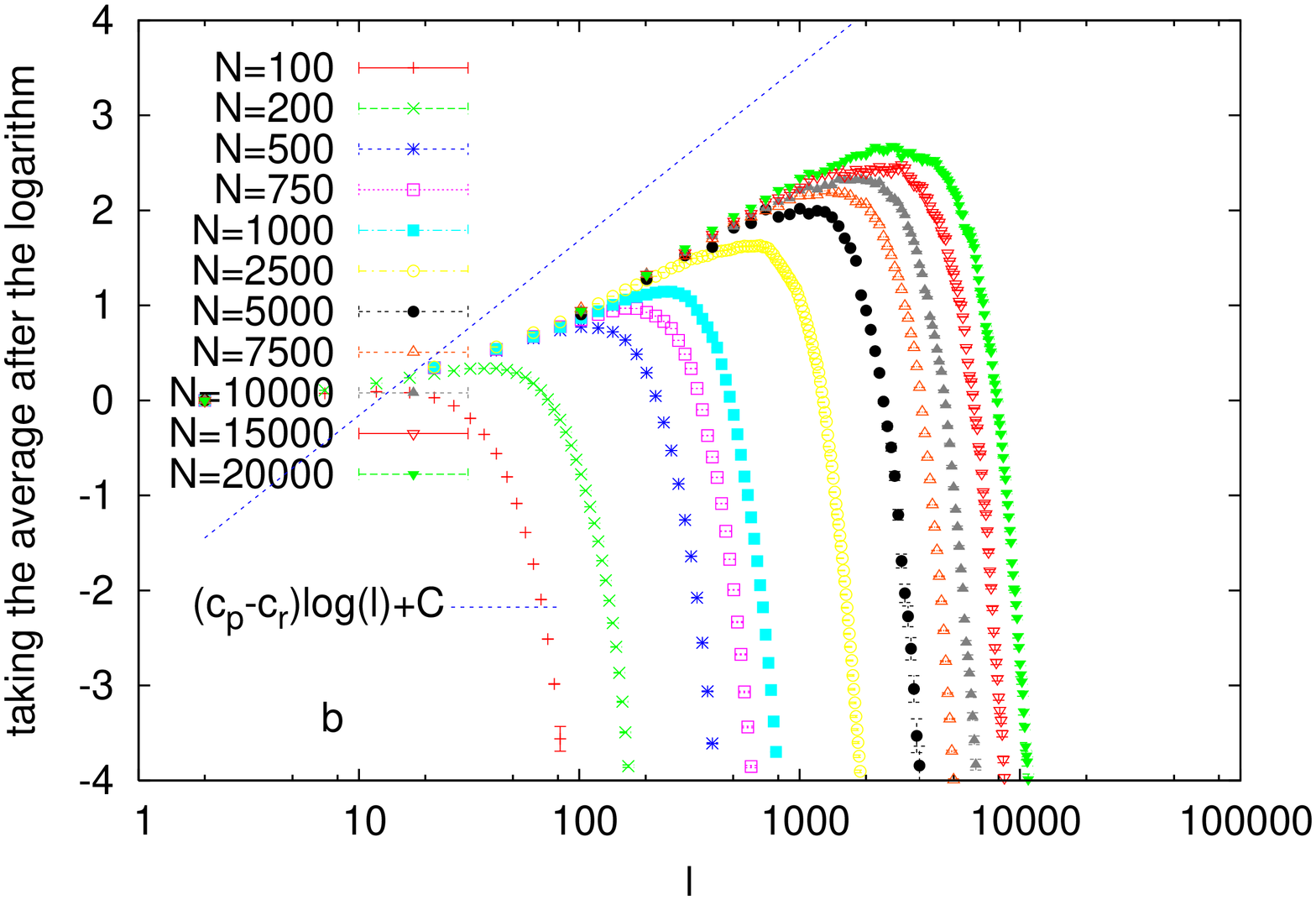,angle=270,width=8cm}
\caption{a) Plots of 
$\log \overline{\overline{P'( \{ \varepsilon_i \},N,T_c,l)}}$ as functions of 
$l$ at $T=T_c \simeq 1.1$, for the different chain lengths considered. 
b) Plots of $\overline{\overline{\log P'( \{ \varepsilon_i \},N,T_c,l)}}$, 
as functions of $l$ at $T=T_c \simeq 1.1$, for the different chain lengths 
considered. For both a) and b) the averages 
are performed by taking into account the pseudo-critical 
temperature, following Eq.~(\ref{newaverages}). 
All curves are shifted arbitrarily, 
setting them to zero for $l=2$. The expected asymptotic limit 
behaviour $\propto (c_p-c_r) \log l+C$, with $c_p-c_r=0.8$ ({\em i.e.} 
$c_r=1.35$), 
is also plotted, in both panels, as a dotted line.}
\label{rev_new_fig12}
\end{center}
\end{figure}

The importance of finite size effects is further highlighted in 
[Fig. \ref{rev_new_fig10}]. This Figure allows to compare 
the behaviour of $1/\xi_2(N,T)$ ([Fig. \ref{rev_new_fig10}a]; as obtained from 
Eq.~(\ref{cr2xi2})), with that of $1/\xi_1(N,T)$ ([Fig. \ref{rev_new_fig10}b]; 
as obtained from Eq.~(\ref{cr1xi1})), with the plots 
clearly showing that the two quantities approach the asymptotic limit 
$\propto (T_c-T)^{\nu_r}$ with $\nu_r \simeq 2.9$ from opposite sides. In 
particular, for short chain lengths $N < N^*$, it appears
that the value of $1/\xi_2(T,N)$ is definitely different 
from zero, on the whole $T$-range considered. It can be also 
observed that, in the same short chain length regime, for the 
average performed after taking the logarithm, negative (physically meaningless)
values are obtained for $\xi_1(T,N)$ on a large part of 
the $T$-range (bearing in mind that in this case it appears necessary to 
insert also a quadratic term in $l$ in order to perform the data fit). 
Even though the data here do not allow a more quantitative
analysis, these results further imply that below the crossover
({\em i.e.} for chain 
lengths $N<2500$) evidence is observed for the presence of two
correlation lengths (and accordingly for two different critical exponents 
$\nu_{r,2} \neq \nu_{r,1}$).

On the other hand, the behaviours of $\xi_1$ and $\xi_2$ for $N > N^*$ are in 
agreement over a large 
$T$-range, within the error margins, with a correlation length exponent 
$\nu_r \simeq 2.9$, as clearly shown
in [Fig. \ref{rev_new_fig11}] displaying the 
plots of $1/\xi_2(N,T)$ and $1/\xi_1(N,T)$ for the three largest sizes 
considered. The fact that the two quantities converge towards the same 
behaviour 
is particularly meaningful: these observables are very
demanding to be measured, and moreover they are the ones expected to be the 
most affected by corrections to scaling. Therefore, the result confirms
that for $N \sim 10^4$ the asymptotic regime of the model considered is 
definitely reached.

Finally, for the 
different chain lengths, 
[Fig. \ref{rev_new_fig12}] displays 
plots of $\log \overline{\overline{P'( \{ \varepsilon_i \},N,T_c,l)}}$ and 
$\overline{\overline{\log P'( \{ \varepsilon_i \},N,T_c,l)}}$ with the averages 
taking into account the  
pseudo-critical temperature, following Eq.~(\ref{newaverages}). More precisely, 
in this case, for any given sequence, the loop-length probability distribution 
which contributes to the average is the one evaluated at the
$T_c( \{ \varepsilon_i \}, N)$, where the 
susceptibility of the sequence 
reaches its absolute maximum. Upon comparing these plots to those 
of $\log {\overline{P'( \{ \varepsilon_i \},N,T_c,l)}}$ and 
${\overline{\log P'( \{ \varepsilon_i \},N,T_c,l)}}$ in
\cite{CoYe} (respectively  [Fig. 6] and  [Fig. 9], in this reference),   
it becomes once more clear that the analysis in terms 
of pseudo-critical temperatures 
allows to reduce the importance of finite size corrections to scaling. 
This can be seen first of all in the fact that here data corresponding to 
average after taking the logarithm display a behaviour more similar to the one 
in which the logarithm is taken after the average. Moreover both  
quantities approach qualitatively more rapidly the asymptotic behaviour 
$\propto (c_p - c_r) \log l + C$, with $c_r \simeq 1.35$, in the present case.

\subsection{Dependence of the crossover chain length on the model parameters}
\label{discussion2_two}
On general grounds, the rounding of the transition due to disorder in the 
present model should be mainly attributable to the presence of  
{\em rare regions} in the sequences (as expected in particular from the 
theoretical results \cite{AiWe}), or otherwise stated to the importance of the 
{\em atypical events} \cite{GiTo1,GiTo2,Gi}. Accordingly, we attempt to 
quantify roughly 
their contribution in the finite size behaviour of the system, to 
better understand the way in which relevance of disorder
becomes manifest, starting from the finite size level.
This approach relies on the numerical evaluation here of 
$N^*$ for the particular model considered, as well as on 
the previously proposed very simplified phenomenological picture \cite{CoYe}.

Qualitatively speaking, the approach is expected to be meaningful in the 
present case with $c_p>2$, corresponding to a first order transition in the 
pure model. Indeed, from the finite size
behaviour point of view, this model appears to be characterized in general by 
a slow approach to the asymptotic regime, and the evidence for the effect 
of disorder (leading to the predicted smooth transition 
described by a single correlation length) appears to be 
related to the quite sudden appearance, at $N \sim N^*$, 
of enhanced (with respect to the $c_p \le 2$ case) multi-step behaviours in 
the order parameter (and accordingly of different peaks in the 
susceptibility) in a significant fraction of the sequences considered.
This feature can then be expected to be related
to the presence in such sequences of {\em large enough rare regions}, which
would then be already occurring, at $N \sim N^*$, with not negligible 
probability.

It is worth recalling that in \cite{CoYe} we showed notably 
that the relevant quantity 
from this point of view is expected to be the adimensional ratio 
between the rare region length $L$ ({\em i.e.} the number of consecutive
base pairs of the same kind) and the 
value of the parameter:
\begin{equation}
x=\frac{R}{R-1} \frac{1}{\log \mu},
\end{equation}
in the particular model {\em \`a la} PS with $c_p= 2.15$
considered.
This quantity is therefore to be interpreted as a measure of the 
{\em effective} disorder strength, linking the concept of large enough 
rare region to the values of $R=\varepsilon_{GC}/\varepsilon_{AT}$ and 
$\log \mu$ in the model. 

Accordingly, we are led to the combinatorial problem of evaluating the
probability ${\cal P}(N,L)$ to observe, in a sequence of total length
$N$, a subsequence of consecutive base pairs of the same kind of length at 
least $L$. Such evaluation could appear to be rather simple. However even the 
obtention of the exact solution for the 
binomial quenched disordered variables distribution given by 
Eq.~(\ref{disorderprobability}) appears to be far from trivial, and we resort 
here to the approximation: 
\begin{equation}
{\cal P}(N,L) \sim (N-L+1) \: 2^{-L}. 
\label{pnl}
\end{equation}
This approximation is exact for $L=N$, but clearly invalid for small 
$L$ values, leading to probabilities higher than 1. Nonetheless, 
it can be considered that this approximation provides 
a reasonable basis for the present analysis, as we are mainly interested to 
capture the 
order of magnitude of the quantity (we checked numerically, 
by exact computations, that this is roughly the case, in a large $L$ range, 
already for small $N$ values).

As already outlined above, at the crossover it is expected to find large enough 
rare regions with a non-negligible probability. Here we set 
this probability to 
the value $0.5$, based on the observation that the order parameter 
displays multi-step behaviour at $N \sim N^*$ ({\em i.e.} multi-peak behaviour 
of the susceptibility) in a fraction of order $o(1)$ of the 
sequences considered. Accordingly, by 
applying Eq.~(\ref{pnl}) with ${\cal P}(N^*,L^*) =0.5$ 
(hence finding numerically the solution $L^*(N^*)$ of
$(N^*-L^*+1)2^{-L^*}=0.5$), we get an estimation 
of the crossover rare region length $L^* \sim 14$, in correspondence with the 
crossover chain length $N^* \sim 10^3 \div 10^4$ which characterizes
the present model. 
Assuming that the phenomenological picture 
does indeed capture the basic physics
of the problem, and recalling
that one has $x=x_{CY} \simeq 1.3$ in the present case, it is now in general 
easy to 
predict the behaviours of $L^{*}(x)$ and of 
the corresponding crossover chain 
length $N^{*}(x)$ as functions of the parameter $x$ in models {\em \`a la} PS 
for DNA denaturation transition with $c_p=2.15$. 
In fact, in order for the effect of disorder to be equally manifest
in different models, it is expected to be necessary for the underlying ratio 
values $L^{*}(x)/x$ (of the
crossover rare region length to the parameter $x$) to be similar \cite{CoYe}. 
Such
conditions can be expressed as $L^*(x)=L^*(x_{CY}) x/x_{CY}$. By using  
$x_{CY}=1.3$ with $L^*(x_{CY})=14$, it is then possible to obtain $N^*(x)$ by 
solving once again Eq.~(\ref{pnl}) with ${\cal P}(N^*,L^*) =0.5$.

In this way, for the PS model with $c_p=2.15$ in 
\cite{GaMo,MoGa}, we find extremely large $N^*$ values of order 
$10^{30} \div 10^{50}$. Indeed, the link energies and the connectivity constant
used in \cite{GaMo,MoGa} (which we recalled in 
[Tab. 3]) lead to $x=x_{GM}\simeq 16$. Taking also into account 
the somehow different law for the allowed
coupling adopted in that study, the effective value of the parameter $x_{GM}$ 
is in fact expected to be still larger. 
Even though relying on rough evaluations, it appears  
therefore reasonable 
to consider that with the conditions in \cite{GaMo,MoGa} it would be
impossible, in practice, to reach the asymptotic regime. 
In such context, the PS model with $c_p=2.15$ in 
\cite{GaMo,MoGa} is in fact indeed expected to behave in agreement with a 
pseudo first order 
transition, ruled by two correlation lengths (corresponding to typical and 
average quantities; namely with $\nu_{r,1}=\nu_p=1$ and $\nu_{r,2}=2$), 
independently from the observable considered and even well beyond the 
(already extremely large) 
chain lengths studied
(up to $N = 2 \cdot 10^6$).  

Thus, in the context of the analysis above, it appears possible to reconcile 
the two different pictures 
emerging from the previous numerical studies \cite{Co,CoYe,GaMo,MoGa}. 
Further, in such context, 
the scenario associated with the pseudo first 
order transition 
appears meaningful to describe the finite-$N$ behaviour
in the presence of {\em weak} disorder.
In this case, it would be moreover interesting
to clarify the importance of boundary conditions, {\em i.e.} whether 
the second correlation length is better interpreted as the one 
related to the free-end distance or to an average loop length, with in fact the
results here concerning the loop-length probability distribution favouring the 
second alternative. 
 
Finally, it is worth noting that the $x_{CY} \simeq 1.3$  
of the present model, which pertains to the region associated with large 
finite size effects (yet possible to study), is  
definitely closer to the still
{\em smaller} $x_{exp} \simeq 0.2$ associated with the values  
of the coupling energies and of $\log \mu$ typically 
adopted for comparisons with experimental melting curves.
In detail, the value of $R$ adopted for comparison
with experiments is essentially the same ($R\simeq 1.1$) than the one 
in \cite{GaMo,MoGa}, with, however, underlying link energies more than
an order of magnitude smaller and a $\log \mu$ value more than an order
of magnitude larger. The simple approximation of $P(L,N)$ 
considered may not be sufficient for precise predictions of 
$L^*_{exp}$ and $N^*_{exp}$ values. However, it seems reasonable to expect
crossover rare region length of the order of a few base pairs,
and a corresponding relatively small crossover chain length. From this point
of view, the present results on the studied PS model with $c_p=2.15$ suggest 
that both self-avoidance and disorder could play a key role in experimental DNA 
behaviour. 

Nevertheless, apart from the possible effect
of neglecting correlations in the sequences, which are well known to be present
in DNA molecules, a throughout understanding of this phenomenon would also 
require to better clarify the role of the cooperativity 
factor $\sigma_0$ in PS models.
Indeed, in order to reproduce correctly the experimental 
curves, usually very small values, $\sigma_0 \sim 10^{-4} \div 10^{-5}$, are 
adopted for 
this parameter (see in particular \cite{BlCa}). It would be also meaningful to
further characterize the influence of this parameter, of non-universal 
character, on the finite size behaviour of the 
system (corresponding roughly to the
introduction of an additional correlation length $\xi_{\sigma_0} \sim 
1/\sigma_0$).

\section{Conclusions}
\label{ending}
We performed an extensive numerical analysis of a disordered PS 
model with $c_p=2.15$, hence with the loop-length probability distribution 
exponent of the pure system taking completely 
into account self-avoidance. The analyses 
are following two main directions: i) a finite 
size scaling study in terms of appropriately defined pseudo-critical
temperatures and ii) an attempt to characterize the peculiar finite 
size behaviour of the model.

Completing previous numerical 
results \cite{Co,CoYe}, and in agreement with the 
predictions based on a probabilistic mathematical approach 
\cite{GiTo1,GiTo2,Gi,To},
the present work notably provides evidence that the 
thermodynamic limit behaviour of the model is coherent with the picture 
describing random ferromagnets \cite{WiDo,AhHa}: a smooth
transition described by a single correlation length (with the refined
estimation here for the corresponding critical exponent 
$\nu_r=2.9 \pm 0.4$).

The pseudo-critical temperature itself, that we define
for taking appropriately into account 
the possible presence of multiple steps in the order parameter for a given
sequence (checking its compatibility with a different definition suggested in 
\cite{GaMo,MoGa} for the same model) appears to be a particularly 
interesting observable to study. 
In detail, we find that the 
mean value and the fluctuations of this pseudo-critical temperature
agree well with the expected scaling laws in the
typical scenario corresponding to disorder relevance. And accordingly, the
two quantities appear to be ruled by the 
same critical exponent (the one corresponding to the correlation length),
on the whole $N$-range considered, without displaying the strong corrections 
to scaling observed in other quantities. 

On this basis, 
it is also interesting to notice that the 
refined estimation of $\nu_r$ obtained with this analysis 
is further compatible with the estimation 
$\nu_r \simeq 2.7$ given in \cite{MoGa}, for a PS model with the different 
$c_p=1.75$. Therefore, the results here could 
also support the hypothesis
that the random PS model critical behaviour is 
independent on the $c_p$ value, as soon as the pure PS model undergoes a 
transition characterized by a diverging specific heat.

From this point of view, it can be nevertheless noticed that we  
focused here on the simplest possible scaling picture, 
therefore that the present study does 
not rule out the possibility that the complete description of the random 
system could involve more than a single independent critical exponent.
In such case, it could be possible to get
a scaling law obeyed by the order parameter and
a better scaling of the loop-length probability distribution at 
$T \neq T_c$. However such possibility, whose exploration would represent a 
new subject, would clearly not impact the present result on $\nu_r$, and it is 
not expected to impact neither
the present results on the characterization of 
the finite size behaviour.

Moreover, in any event, the observation that in the present model the 
order 
parameter behaves as the energy, and the susceptibility as the specific heat, 
could be relevant only for the disordered regime considered. More 
specifically, we did not investigate here the {\em strong} disorder regime of 
the model. Indeed, for $\varepsilon_{GC} >> \varepsilon_{AT}$, the observed 
behaviour could be more complex, since models {\em \`a la} PS are expected 
to display an additional singularity of the Kosterlitz-Thouless kind 
\cite{TaCh,KaMu,DeRe}.

The present analysis also shows qualitatively that the finite size 
corrections are in general smaller when quantities are averaged taking into
account the pseudo-critical temperatures. Finally, based on the study of 
an appropriately defined non self-averageness parameter, evidence is provided 
that the susceptibility is a strongly non self-averaging quantity at the 
critical point, with its detailed behaviour further allowing to get an 
alternative evaluation of both $T_c$ and $\nu_r$, compatible with our 
previous ones.

In this background, mainly in the second part of the 
work, we seek a better understanding of finite size effects. 
The data for the maximum of the susceptibility, 
and even more so the detailed study of the behaviour of the loop-length 
probability distribution, give weight to the existence of a 
crossover chain length $N^*$ below which one could not rule out 
the picture proposed in \cite{GaMo,MoGa}. This alternative scenario, 
the pseudo first order phase transition one, is indeed found to capture 
the behaviour of a different PS model with the same $c_p=2.15$ up to the
extremely large sizes of $N=2 \cdot 10^6$.
In detail, we present a qualitative evaluation from below of
$N^* \ge 1000$ from data on the pseudo-$T_c$, and a more quantitative
evaluation from above of $N^* \le 2500 \div 5000$ from data on the 
loop-length probability distribution.

We also provide a tentative quantification of the dependence of $N^*$ on the 
parameters of the model, with notably an evaluation of the 
crossover length behaviour which could allow to reconcile the results here with 
those in \cite{GaMo,MoGa}: 
with the parameters chosen for the model in \cite{GaMo,MoGa} it could be not 
possible to reach the asymptotic regime in practice. In this context, we also 
find that the 
crossover chain length obtained for realistic parameter values (as used in 
experimental settings) should be definitely smaller than
the present $N^*$ and, accordingly, our conclusions, concerning both the 
importance of self-avoidance and the relevance of disorder, should be also 
important for better modeling experimental DNA denaturation. 

It is finally also interesting to stress that these extensive studies 
of the finite size behaviour, particularly in 
the case of PS models for DNA denaturation transitions, were made possible 
thanks to the recursive equations for the partition functions
(notably within the SIMEX scheme), reducing 
essentially the time complexity of the problem to $\ln(N)N$.  

In conclusion, it is possible that
crossover effects, such as those described here, 
could be relevant in a 
larger class of disordered systems (at 
least in the behaviour of 
certain observables), allowing a better understanding of the finite 
size behaviour, and in particular of the way in which relevance of disorder 
becomes manifest when approaching the thermodynamic limit. 
In this sense the results and treatments 
here could provide interesting insights for the exploration of such 
disordered systems, on more general grounds. 
On the biological side, the results here should also contribute to
a better understanding of the roles played by disorder and self-avoidance in 
experimental DNA denaturation, with notably a quantitative description of the 
finite size effect particularly important in this context. 

\section*{Acknowledgments}
B.C. would like to acknowledge the {\em Unit\'e} BIS of the Pasteur Institut 
and the {\em CERES-ERTI} of the ENS in Paris, for post-docs stays during which 
the work was initiated. She is moreover grateful to
Alberto and Enrico Bersani for many encouragements and clarifying discussions.
We thank Nicoletta Cancrini, Giorgio Parisi, Andrea Pelissetto and Allegra
Via for comments, as well as two anonymous referees for insightful remarks.
E.Y. acknowledges (non-financial) support from the ANR Mod\`eles Num\'eriques 
(ANR-11-MONU-0020: ProbAlg).


\begin{thebibliography}{99}

\bibitem{GiTo1}
G. Giacomin and F.L. Toninelli, {\em Smoothing of depinning transitions for 
directed polymers with quenched disorder}, 2006,
{\em Phys. Rev. Lett.} {\bf 96}, 060702.

\bibitem{GiTo2}
G. Giacomin and F.L. Toninelli, {\em Smoothing effect of quenched disorder on 
polymer depinning transitions}, 2006, 
{\em Commun. Math. Phys.}  {\bf 266}, 1. 

\bibitem{Gi}
G. Giacomin,
{\em Random Polymer Models}, 2007 (Imperial College Press, London). 

\bibitem{CoYe}
B. Coluzzi and E. Yeramian, {\em Numerical evidence for relevance of disorder 
in a Poland-Scheraga DNA denaturation model with self-avoidance: scaling 
behavior of average quantities}, 2007, {\em Eur. Phys. J. B} {\bf 56}, 349.

\bibitem{GaMo}
T. Garel and C. Monthus, {\em  Numerical study of the disordered 
Poland–Scheraga model of DNA denaturation}, 2005 {\em J. Stat. Mech.}, P06004.

\bibitem{MoGa}
C. Monthus and T. Garel, {\em Distribution of pseudo-critical temperatures 
and lack of self-averaging in disordered Poland-Scheraga models with different 
loop exponents}, 2005, {\em Eur. Phys. J. B} {\bf 48}, 393. 

\bibitem{To}
F.L. Toninelli, {\em Correlation lengths for random polymer models and for 
some renewal sequences}, 2007,
{\em Electron. J. Probab.} {\bf 12}, 613.  

\bibitem{PoSh1}
D. Poland and H.A. Scheraga,  {\em Phase transitions in one dimension and the 
helix—coil transition in polyamino acids}, 1966, 
{\em J. Chem. Phys.} {\bf 45}, 1456; {\em Occurrence of a phase transition in 
nucleic acid models}, 1966, {\em J. Chem. Phys.} {\bf 45}, 1464; 

\bibitem{PoSh2}
D. Poland and H.A. Scheraga (eds.),
{\em Theory of Helix-Coil Transitions in Biopolymers}, 1970, (Academic,
New York). 

\bibitem{Fi1}
M.E. Fisher, {\em Effect of excluded volume on phase transitions in 
biopolymers}, 1966, {\em J. Chem Phys.} {\bf 45}, 1469.

\bibitem{review}
R.M. Wartell and A.S. Benight, {\em Thermal denaturation of DNA molecules: 
A comparison of theory with experiment}, 1985, {\em Phys. Rep.} {\bf 126}, 67.

\bibitem{Bl} R.D. Blake, {\em et al.}, {\em Statistical mechanical simulation 
of polymeric DNA melting with MELTSIM}, 1999, {\em Bioinformatics},
{\bf 15}, 370.

\bibitem{JoEv} D. Jost and R. Everaers, {\em A unified Poland-Scheraga model 
of oligo- and polynucleotide DNA melting: salt effects and predictive power},
2009, {\em Biophys. J.}  {\bf 96}, 1056.

\bibitem{Ye}
E. Yeramian, {\em Genes and the physics of the DNA double-helix}, 2000,
{\em Gene} {\bf 255}, 139; {\em The physics of DNA and the annotation of the 
Plasmodium falciparum genome}, 2000,
{\em Gene} {\bf 255}, 151.

\bibitem{YeJo}
E. Yeramian and L. Jones, {\em GeneFizz: a web tool to compare genetic 
(coding/non-coding) and physical (helix/coil) segmentations of DNA sequences. 
Gene discovery and evolutionary perspectives}, 2003,
{\em Nucl. Acid Res.} {\bf 31}, 3843.

\bibitem{YeBoLa}
E. Yeramian, S. Bonnefoy and G. Langsley, {\em Physics-based gene 
identification: proof of concept for Plasmodium falciparum}, 2002,
{\em Bioinformatics} {\bf 18}, 190.

\bibitem{CaMaBl}
E. Carlon, M.L. Malki and R. Blossey,
{\em Exons, Introns, and DNA Thermodynamics}, 2005,
{\em Phys. Rev. Lett.} {\bf 94}, 178101.

\bibitem{Fi0}
M.E. Fisher, {\em Walks, walls, wetting, and melting}, 1984, 
{\em J. Stat. Phys.} {\bf 34}, 667.

\bibitem{KaMuPe1}
Y. Kafri, D. Mukamel and L. Peliti,
{\em Why is the DNA Denaturation Transition First Order?}, 2000,
{\em Phys. Rev. Lett.} {\bf 85}, 4988.
A. Hanke and R. Metzler,  {\em Comment}, 2003, 
{\em Phys. Rev. Lett.} {\bf 90}, 159801; 
Y. Kafri, D. Mukamel and L. Peliti, {\em Reply}, 2003,
{\em Phys. Rev. Lett.} {\bf 90} 159802.

\bibitem{KaMuPe2}
Y. Kafri, D. Mukamel and L. Peliti, {\em Melting and unzipping of DNA},
2002, {\em Eur. Phys. J. B} {\bf 27}, 135.

\bibitem{Du}
B. Duplantier, {\em Polymer Network of fixed topology: renormalization, exact 
critical exponent $\gamma$ in two dimensions, and $d=4-\varepsilon$}, 1986, 
{\em Phys. Rev. Lett.} {\bf 57}, 941; {\em Statistical mechanics of polymer 
networks of any topology}, 1989,
{\em J. Stat. Phys.} {\bf 54}, 581.

\bibitem{CaCoGr}
M. S. Causo, B. Coluzzi and P. Grassberger,
{\em Simple model for the DNA denaturation transition}, 2000,
{\em Phys. Rev. E} {\bf 62}, 3958.

\bibitem{CaOrSt}
E. Carlon, E. Orlandini and A.L. Stella,
{\em Roles of stiffness and excluded volume in DNA denaturation}, 2002,
{\em Phys. Rev. Lett.} {\bf 88}, 198101.

\bibitem{BaCaSt}
M. Baiesi, E. Carlon and A.L. Stella, {\em Scaling in DNA unzipping models: 
Denaturated loops and end segments as branches of a block copolymer network},
2002, {\em Phys. Rev. E}, {\bf 66}, 021804.

\bibitem{BaCaKaMuOrSt}
M. Baiesi, {\em et al.}, {\em Interstrand distance distribution of DNA near 
melting}, 2003, {\em Phys. Rev. E}, {\bf 67}, 021911.

\bibitem{KuLi} H. Kunz and R. Livi, {\em DNA denaturation and wetting in the 
presence of disorder}, 2012, {\em Europhys. Lett.} {\bf 99}, 30001.

\bibitem{Ha}
A.B. Harris, {\em Effect of random defects on the critical behaviour of Ising 
models}, 1974, {\em J. Phys. C} {\bf 7}, 1671.

\bibitem{ChChFiSp}
J.T. Chayes, L. Chayes, D.S. Fisher and T. Spencer,
{\em Finite-size scaling and correlation lengths for disordered systems},
1986, {\em Phys. Rev. Lett.} {\bf 57}, 2999.

\bibitem{AiWe} 
M. Aizenman and J. Wehr, {\em Rounding of first-order phase transitions in 
systems with quenched disorder}, 1989,  {\em Phys. Rev. Lett.} {\bf 62},
2503; {\em Erratum}, 1990, {\em Phys. Rev. Lett.} {\bf 64}, 1311.

\bibitem{Be} 
A.N. Berker, {\em Critical behavior induced by quenched disorder}, 1993, 
{\em Physica A} {\bf 194}, 72.

\bibitem{AhHa} 
A. Aharony and A.B. Harris, {\em Absence of Self-Averaging and Universal Fluctuations in Random Systems near Critical Points}, 1996, 
{\em Phys. Rev. Lett.} {\bf 77}, 3700.

\bibitem{Br} 
R. Brout, {\em Statistical Mechanical Theory of a Random Ferromagnetic 
System}, 1959, {\em Phys. Rev.} {\bf 115}, 824.

\bibitem{WiDo}
S. Wiseman and E. Domany, {\em Finite-Size Scaling and Lack of Self-Averaging 
in Critical Disordered Systems}, 1998, {\em Phys. Rev. Lett.} {\bf 81}, 22; 
{\em Self-averaging, distribution 
of pseudocritical temperatures, and finite size scaling in critical 
disordered systems}, 1998, {\em Phys. Rev. E} {\bf 58}, 2938.

\bibitem{Co}
B. Coluzzi, {\em Numerical study of a disordered model for DNA denaturation 
transition}, 2006,
{\em Phys. Rev. E}, {\bf 73} 011911.

\bibitem{Ba}
M.N. Barber, {\em Finite Size Scaling}, in {\em Phase Transitions and
Critical Phenomena}, Vol. 8, 1983, edited by C. Domb and J.L. Lebowitz
(Academic, New York).

\bibitem{SeShTa}
W. Selke, L.N. Shchur, and A.L. Talapov, {\em Monte Carlo Simulations of
Dilute Ising Models}, in {\em Annual Reviews of Computational Physics}, 
1994, edited by D. Stauffer (World Scientific, Singapore).

\bibitem{BiYo}
K. Binder and P. Young, {\em Spin glasses: Experimental facts, theoretical 
concepts, and open questions}, 1986, {\em Rev. Mod. Phys.} {\bf 58}, 801.

\bibitem{ZJ}
J. Zinn-Justin, {\em Quantum Field Theory and Critical Phenomena},  
Fourth Edition, 2002 (Clarendon Press, Oxford).

\bibitem{DFi}
D.S. Fisher, {\em Random transverse field Ising spin chains}, 1992 
{\em Phys. Rev. Lett.} {\bf 69}, 534;
{\em Critical behavior of random transverse-field Ising spin chains},
1995, {\em Phys. Rev. B} {\bf 51}, 6411.

\bibitem{IgLiRiMo}
F. Igl\'oi, Yu-Cheng Lin, H. Rieger, C. Monthus, {\em Finite-size scaling of 
pseudocritical point distributions in the random transverse-field Ising chain},
2007, {\em Phys. Rev. B} {\bf 76}, 064421.

\bibitem{MoGa2}
C. Monthus and T. Garel, {\em Random wetting transition on the Cayley tree: a 
disordered first-order transition with two correlation length exponents},
2009, {\em J. Phys. A: Math. Theor.} {\bf 42}, 165003.

\bibitem{Bietal}
A. Billoire, et al, {\em Finite size scaling analysis of the distribution 
of critical temperatures in spin glasses}, 2011, {\em J. Stat. Mech.}, P10019.

\bibitem{FiFr}
M. Fixman and J.J. Freire, {\em Theory of DNA melting curves}, 1977,
{\em Biopolymers} {\bf 16}, 2693.

\bibitem{Yeetal}
E. Yeramian, {\em et al.}, {\em An optimal formulation of the matrix method 
in statistical mechanics of one-dimensional interacting units: Efficient 
iterative algorithmic procedures}, 1990,
{\em Biopolymers} {\bf 30}, 481.

\bibitem{Ye1}
E. Yeramian, {\em Complexity and tractability. Statistical mechanics of 
helix-coil transitions in circular DNA as a model-problem}, 1994,
{\em Europhys. Lett.} {\bf 25}, 49.

\bibitem{YeCl}
E. Yeramian and P. Claverie, {\em Analysis of multiexponential functions 
without a hypothesis as to the number of components}, 1987, {\em Nature} 
{\bf 326}, 169.

\bibitem{ClDeYe}
P. Claverie, A. Denis and E. Yeramian, {\em The representation of functions 
through the combined use of integral transforms and Pad\'e approximants: 
Pad\'e-Laplace analysis of functions as sums of exponential}, 1989, 
{\em Comput. Phys. Rep.} {\bf 9}, 247. 

\bibitem{YeDe}
E. Yeramian and E. Debonneuil, {\em Probabilistic sequence alignments: 
Realistic models with efficient algorithms}, 2007,
{\em Phys. Rev. Lett.} {\bf 98}, 078101.

\bibitem{BlCa}
R. Blossey and E. Carlon, {\em Reparametrizing the loop entropy weights: 
Effect on DNA melting curves}, 2003,
{\em Phys. Rev. E} {\bf 68}, 061911.

\bibitem{TaCh}
L.-H. Tang and H. Chat\'e, {\em Rare-event induced binding transition of 
heteropolymers}, 2001, {\em Phys. Rev. Lett.} {\bf 86}, 830.

\bibitem{KaMu}
Y. Kafri and D. Mukamel, {\em Griffiths singularities in unbinding of strongly 
disordered polymers}, 2003, {\em Phys. Rev. Lett.} {\bf 91}, 055502.

\bibitem{DeRe}
B. Derrida and M. Retaux, {\em The depinning transition in presence of 
disorder: A toy model}, 2014, {\em J. Stat. Phys.} {\bf 156}, 268.

\end{thebibliography}
\end{document}